\documentclass[final,3p,twocolumn]{elsarticle}
\usepackage{dcolumn}
\usepackage{lineno}
\modulolinenumbers[5]
\usepackage{amssymb}
\usepackage[fleqn]{amsmath}
\usepackage{mathptmx}
\usepackage[colorlinks=true,linkcolor=blue,urlcolor=blue,citecolor=blue]{hyperref}
\biboptions{numbers,sort&compress}

\journal{Journal of Magnetism and Magnetic Materials}
\begin{document}
\begin{frontmatter}
	\title{Density matrix renormalization group approach to the low temperature thermodynamics of correlated 1D fermionic models}
\author[kolkata]{Sudip Kumar Saha\corref{}}
\author[kolkata]{Debasmita Maiti\corref{}}
\author[kolkata]{Manoranjan Kumar\corref{}}
\author[New Jersey]{Zolt\'an G. Soos\corref{}}

\address[kolkata]{S. N. Bose National Centre for Basic Sciences, Block JD, Sector III, Salt Lake, Kolkata - 700106, India}
\address[New Jersey]{Department of Chemistry, Princeton University, Princeton, New Jersey 08544, USA}

\ead{manoranjan.kumar@bose.res.in, soos@princeton.edu}

\date{\today}

\begin{abstract}
The low temperature thermodynamics of correlated 1D fermionic models with spin and charge degrees of freedom is obtained
by exact diagonalization (ED) of small systems and followed by density matrix renormalization group (DMRG) calculations that
target the lowest hundreds of states $\{E(N)\}$ at system size $N$ instead of the ground state. Progressively larger $N$
reaches $T < 0.05t$ in correlated models with electron transfer $t$ between first neighbors and bandwidth $4t$. The size dependence of
the many-fermion basis is explicitly included for arbitrary interactions by scaling the partition function. The remaining size
dependence is then entirely due to the energy spectrum $\{E(N)\}$ of the model. The ED/DMRG method is applied to Hubbard and
extended Hubbard models, both gapped and gapless, with $N_e = N$ or $N/2$ electrons and is validated against exact results for
the magnetic susceptibility $\chi(T)$ and entropy $S(T)$ per site. Some limitations of the method are noted. Special attention is given to the
bond-order-wave phase of the extended Hubbard model with competing interactions and low $T$ thermodynamics sensitive to small gaps.

\end{abstract}


\end{frontmatter}


\section{\label{sec:intro}Introduction}

Quantum cell models have a finite basis that diverges in the thermodynamic limit. The full energy spectrum $\{E(N)\}$ is directly accessible for 
systems of tens of interacting spins or fermions. The size dependence of $1$D models with open or periodic boundary conditions has long been 
studied in such contexts as trans polyenes with $N/2$ double bonds or as $1/N$ extrapolations to the thermodynamic limit. Bonner and Fisher~\cite{bonner64} 
inferred the thermodynamics of the linear Heisenberg antiferromagnet (HAF), the spin-$1/2$ chain with isotropic nearest-neighbor 
exchange $J > 0$, by exact diagonalization (ED) up to $N = 12$, judicious analysis of size dependencies and reference to exact $T = 0$ results. 
The same approach has been applied to Hubbard and extended Hubbard models, to the Pariser-Parr-Pople model for $\pi$-electrons in conjugated hydrocarbons, 
and to $1$D models of molecular crystals with metal-insulator or neutral-ionic transitions. Fermionic models have charge and spin degrees of freedom and typically 
have one or two sites per unit cell. We develop in this paper a general numerical approach to the low-$T$ thermodynamics of correlated $1$D fermionic models.

The basic idea is that the full spectrum $\{E(N)\}$ of large systems should never be needed. ED of small systems returns the thermodynamics 
for $T > T(N)$ when $N$ exceeds the correlation length. Convergence at high $T$ follows directly from the size independence of per-site quantities.
Density matrix renormalization group (DMRG) calculations then give the low-energy states $E(N)$ of larger systems. We seek hundreds of states $R(N)$ 
that minimize finite-size effects and reach the thermodynamic limit at $T(N)$ at system size $N$. Thermodynamics at progressively lower $T(N)$ is 
obtained on increasing $N$. The explicit dependence on system size provides useful guidance for convergence or extrapolation. Aside from targeting $R(N)$ 
states, the ED/DMRG approach follows conventional DMRG~\cite{white-prb93}, now a powerful and well-established method~\cite{schollwock2005,karen2006}, 
for the quantum ($T = 0$ K) phases and lowest excitations of $1$D models.

We mention other approaches to the thermodynamics of correlated $1$D models. 
The transfer matrix renormalization group (TMRG) involves expressing the partition function in terms of trace of transfer matrix (TM) products and calculating free
energy and other thermodynamic quantities from the maximum eigenvalue of the TM~\cite{nishino1995,peschel99}. TM of a certain size gives the thermodynamic limit at a certain $T$ without the
need of any extrapolation in system size. Lower $T$ is reached with progressively larger system size $N$ and, as a result, has larger truncation error. The lowest 
$T$ accessible for the frustrated $J_1-J_2$ model with truncation error $< 10^{-4}$ is $T \sim 0.01 J_1$ ~\cite{xiang2006}, and lower $T$ becomes unreliable.  
Fermionic models have similar errors~\cite{sirker2007} at low $T$. 
Numerical considerations limit the lowest attainable $T$ at present, which is the 
regime we focus on. Monte Carlo methods, not limited to $1$D, are not applicable to models with frustrated interactions as they give rise to
sign problems~\cite{sandvik2010}. These methods work very well for half-filled Hubbard models except at very large $U$ where
difficulty arises due to frustration. Exact results rely on the 
Bethe ansatz for models with one spin or site per unit cell~\cite{bethe31,lieb68,johnston2000}. The exotic quantum phases of spin-$1/2$ chains with frustrated exchange $J_1$ and $J_2$ between 
first and second neighbors have been intensively studied theoretically using field theory~\cite{chubukov1991,hikihara2008}. The methods mentioned above 
have been mainly applied to spin chains, which also served as the initial tests~\cite{sudip19} and applications~\cite{sudipsp2020, sudipspjmmm20} of ED/DMRG.


Methods for correlated models can readily be tested against the band or noninteracting limit. We consider $1$D fermionic systems of $N$ sites with 
periodic boundary conditions, $N_e$ electrons or holes and electron transfer $t$ with spin conservation between first neighbors. The noninteracting 
limit is a tight-binding band with single-particle energies $\varepsilon(k) = -2t\cos{k}$, bandwidth $4t$ and wavevectors $-\pi < k = 2\pi m/N \le \pi$ in the first 
Brillouin zone. The Hubbard model has on-site repulsion $U > 0$ while the extended Hubbard model (EHM) also has spin 
independent interaction V between neighbors. The same analysis holds for spin-independent interactions $V_r$ of any range. The models 
conserve the total spin $S \le N_e/2$ and its $z$ component $S^Z$.

The EHM Hamiltonian with sites $r$ and $t = 1$ as the unit of energy is
 \begin{equation}
\begin{aligned}
         H(U,V) &= -\sum_{r,\sigma} {({c}_{r,\sigma}^{\dagger} {c}_{r+1,\sigma} +h.c. )} \\ 
         &   +U \sum_{r} {{c}_{r,\alpha}^\dagger 
        {c}_{r,\beta}^\dagger {c}_{r,\beta} {c}_{r,\alpha}} 
          +V \sum_{r} {{n}_{r} {n}_{r+1}}.
 \label{eq:EHM_ham}
\end{aligned}
 \end{equation}
The operators $c_{r,\sigma}^{\dagger} (c_{r,\sigma})$ create (annihilate) an electron with spin $\sigma$ at site $r$, $h.c.$
is the Hermitian conjugate and $n_r = \sum_\sigma c_{r,\sigma}^\dagger c_{r,\sigma}$ is the number operator. The half-filled Hubbard model
with $n = N_e/N = 1$ and $V = 0$ has been of special interest mathematically. Using the Bethe ansatz, Lieb and Wu~\cite{lieb68} proved
that $U > 0$ opens a gap for conduction and generates a paramagnetic Mott insulator. Takahashi~\cite{takahashi70} obtained the magnetic
susceptibility at $T = 0$. More recently, J\"uttner, Kl\"umper and Suzuki~\cite{juttner98} obtained the exact
thermodynamics and elementary excitations at arbitrary filling, and they related their results to other models to which the Bethe ansatz
is applicable. Glocke  et al. reported~\cite{sirker2007} a comprehensive TMRG study of the thermodynamics of Eq.~\ref{eq:EHM_ham} and found accurate 
spin and charge susceptibilities down to $T/t \sim 0.05$. 

The ground state of the half-filled band with $U = V = 0$ in Eq.~\ref{eq:EHM_ham} has doubly occupied states $\varepsilon(k)$ up 
to the Fermi wavevector $k_F = \pm \pi/2$. The magnetic susceptibility is
\begin{equation}
  \chi(T) = \frac {\beta} {4\pi} \int_{0}^{\pi/2} \frac {dk}{{\cosh^2}(\beta \cos{k})}  
 \label{eq:pauli_paramag}
 \end{equation}
with $\beta=1/T$ in reduced ($T/t$) units. The band limit poses a natural and challenging test of  methods based on finite systems.
At issue is the extreme size dependence of the $T \sim 0$ thermodynamics that follows from the H\"uckel $4p$, $4p + 2$ rule. The ground state
is 6-fold degenerate (a triplet and three singlets) when $N_e = 4p$. There are two electrons in the degenerate states at $k_F = \pm \pi/2$
when $N = N_e$ and $\chi(T,N)$ diverges as $1/NT$. The singlet ground state is nondegenerate when $N_e = 4p + 2$, all states up
to $k_F$ are doubly occupied and $\chi(0,N) = 0$.

\begin{figure}[t]
\includegraphics[width=\columnwidth]{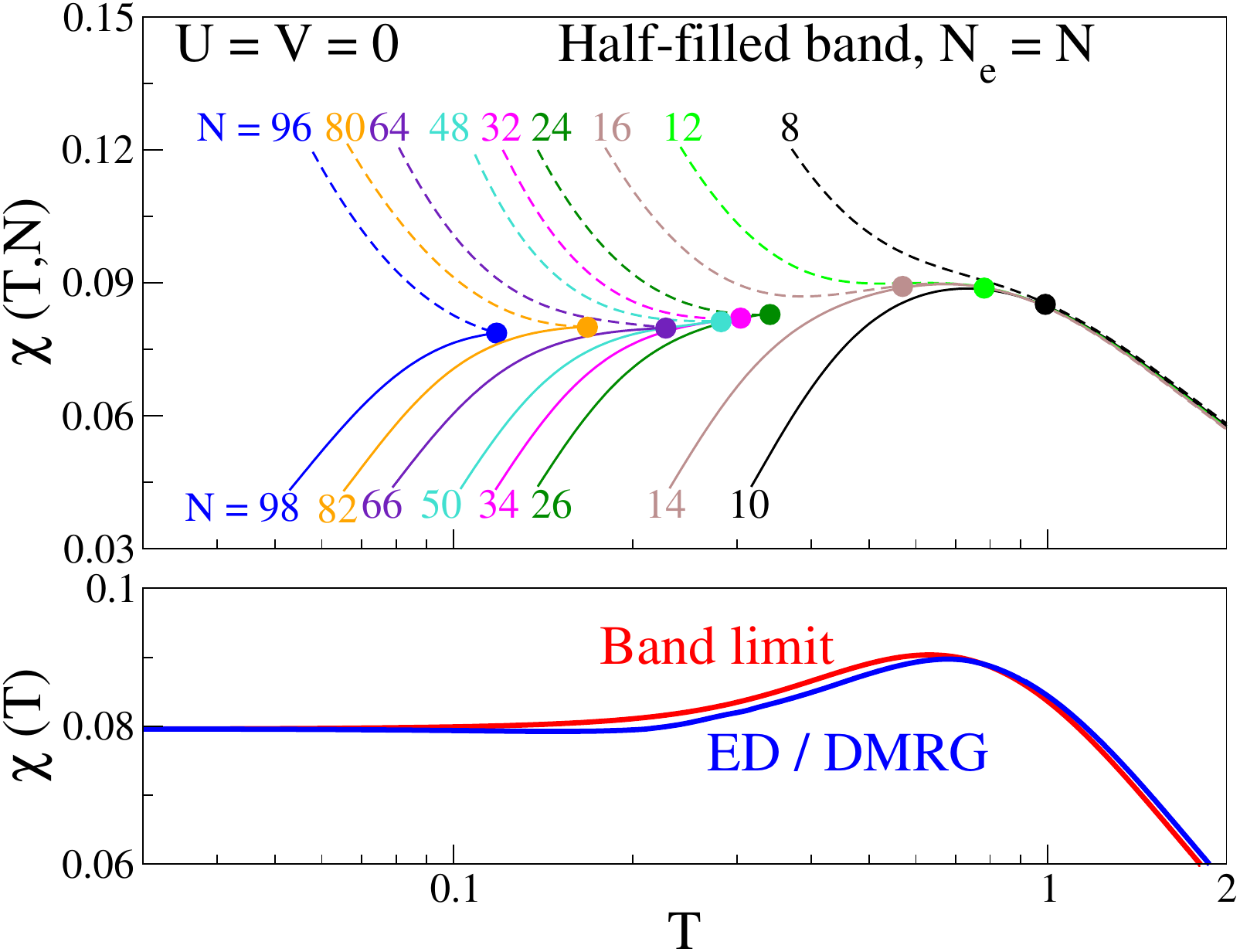}
\caption{\label{fig1}
        Upper panel: Molar magnetic susceptibility $\chi(T,N)$ of systems with $N_e = N$ and $U = V = 0$ in Eq.~\ref{eq:EHM_ham}, exact for $N \le 14$. DMRG for larger $N$ is
shown up to the crossing or merging points $T_N$ at system size $N = 4p$, $4p + 2$.
Lower panel: The band limit is Eq.~\ref{eq:pauli_paramag}; ED/DMRG is through the points
$T_N$ and extrapolated to $T = 0$.}
 \end{figure}

The upper panel of Fig.~\ref{fig1} shows $\chi(T,N)$ at the indicated system sizes,
exact up to $N$ = 14. Larger $N$ is based on DMRG and a truncated energy spectrum 
discussed later. The $\chi(T,N)$ curves merge or cross from above and below for $N = 4p$
and $4p + 2$, respectively, at $T_N$ shown as points. The points clearly require only the low-energy part of the spectrum $\{E(N)\}$. The blue line in the lower panel is $\chi(T,14)$ at high $T$, connects the points $T_N$ at low $T$ and is extrapolated to $T = 0$. The red line is the thermodynamics limit, Eq.~\ref{eq:pauli_paramag}, with $\chi(0) = 1/4\pi$. The ED/DMRG line deviates from the band result at intermediate and high $T$ for reasons 
discussed in Section~\ref{sec2}. The method is approximate for fermionic systems, in contrast to spin chains~\cite{sudip19} where $\chi(T,N)$ converges to $\chi(T)$ exactly
at high $T$.

Spin degrees of freedom dominate the low $T$ thermodynamics of half-filled models with $U \geq 4$ or $U - V \geq 4$ in Eq.~\ref{eq:EHM_ham}. 
Klein and Seitz~\cite{klein73} derived the spin Hamiltonian of the half-filled Hubbard model in powers of $(t/U)^2$, as required for 
virtual transfers between spin states with $n_r = 1$ at all sites. The first two terms are exchanges $J_1$ and $J_2$ between first and second neighbors,
\begin{equation}
{J}_{1} =\frac{{4t}^{2}}{U} -\frac{16{t}^{4}} {{U}^{3}},   \qquad   {J}_{2} = \frac{4{t}^{4}} {{U}^{3}}.
 \label{eq:fermion_j1j2}
 \end{equation}
Von Dongen~\cite{van94} obtained the same exchanges for the EHM with $U - V$ instead of $U$ in the denominator of 
the $t^2$ term and $(U - V)^3/(1 + V/U)$ instead of $U^3$ in the denominator of the $t^4$ term. The Hamiltonian 
with periodic boundary conditions and spins $S_r = 1/2$ is
\begin{equation}
	H(J_1,J_2) = J_1\sum_{r} \vec{S}_r \cdot \vec{S}_{r+1} + J_2 \sum_{r} \vec{S}_{r} \cdot \vec{S}_{r+2}.
\label{eq:j1j2}
\end{equation}
Spin-1/2 chains describe the low $T$ thermodynamics of half-filled fermionic models when interactions exceed the bandwidth. The smaller Hilbert space of $2^N$ spin states is of course very advantageous numerically.

The paper is organized as follows. The method developed in Section~\ref{sec2} begins
with the scaled partition function that accounts for the size dependence of the
many-fermion basis. The DMRG procedure is then summarized for targeting hundreds of correlated states $E_p(N)$ 
in sectors $S^Z \le N_e/2$ at system size $N$ and compared to
excitations of noninteracting fermions. Finally, energy cutoffs with $E_p(N) \le E_C(N)$ 
are introduced to obtain the low $T$ thermodynamics in a narrow range at system size
$N$. Section~\ref{sec3} discusses representative applications of Eq.~\ref{eq:EHM_ham} that include half-filled Hubbard models with 
(a) $U \le 4$, the bandwidth; (b) $U > 4$ and spin-charge separation with low $T$ thermodynamics given by Eq.~\ref{eq:j1j2}; 
(c) Quarter-filled models; (d) Half-filled EHM with $U = 4$ and $V \sim 2$ in the range of the bond-order-wave (BOW) quantum phase. 
ED/DMRG results, primarily for $\chi(T)$ and the entropy $S(T)$ per site, are related to exact and TMRG results. Section~\ref{sec4} 
contains brief comments about the method.

\section{\label{sec2} Methods}

\subsection{\label{sec2a} Scaled partition function}

Fermionic models with $N$ sites and $N_e$ electrons have a large but finite number of states $W(N,N_e)$. Since the Boltzmann factors are unity in the high $T$ limit, the
entropy density or entropy per site is $N^{-1}\ln{W(N,N_e)}$ and must be a continuous
function of $n = N_e/N \le 2$ in the thermodynamic limit. Fermionic systems differ 
in this respect from spin-S systems with $W(N,S) = (2S + 1)^N$ and $N^{-1}\ln{W(N,S)}$
= (2S + 1) for any integer $N$. To obtain the proper high $T$ limit, we introduce 
below a scale factor $\lambda(N,n)$ that returns for finite $N$ the thermodynamic limit at $n = N_e/N$.

Distributions of $N_e$ fermions on $N$ sites lead to $q \leq N_e/2$ doubly occupied sites with $n_r = 2$, $N_e - 2q$ singly occupied sites 
with $n_r = 1$ and spin $\alpha$ or $\beta$, and $N - N_e + q$ unoccupied sites with $n_r = 0$. The dimension of the many fermion basis is
\begin{equation}
W(N, {N}_{e}) = \sum_{q=0}^{{N}_{e}/2} \frac {{N!}\, {2}^{{N}_{e}-2q}} {{q!}\, (N- {N}_{e} +q )!\, ({N}_{e}-2q)!} . 
\label{eq:finitebasis}
\end{equation}
The probability of $q$ doubly occupied sites is $W(q,N,N_e)/ W(N,N_e)$, and the most probable distribution becomes exact in the thermodynamic 
limit. We treat $q/N$ as continuous, find the maximum of $\ln{W(q,N,n)}$ at $n = N_e/N$ with respect to $q$ and obtain $q_{mp}(n)/N = n^2/4$. 
With this value for $q$, the thermodynamic limit of $N^{-1}\ln{W(N,N_e)}$ is
\begin{equation}
	\begin{aligned}
		{N}^{-1} \ln{W(N,n)} & \rightarrow \ln{\left(4{n}^{-n}{(2-n)}^{n-2}\right)} \\ 
		& \equiv \ln{g(n)}. 
\label{eq:densityhighTthermo}
	\end{aligned}
\end{equation}
The straightforward interpretation is that there are $nN/2$ sites with either spin $\alpha$ or $\beta$. The density of doubly occupied 
sites is $n^2/4$; the density of unoccupied sites is $(1 - n/2)^2$; the remaining $n(1 - n/2)$ sites are singly occupied. 
We define the scale factor $\lambda(N,n)$ as
\begin{equation}
\lambda \left(N,n \right) W \left(N,{N}_{e}\right) = {g(n)}^{N}
\label{eq:scale_factor}
\end{equation}
Scaling increases $W(N,N_e)$ such that it matches the thermodynamics limit. We have $g(1) = 4$ at half filling, $g(1/2) = g(3/2) = 16/3^{3/2}$ 
and $g(n) \to 0$ for $n \to 0$ or $2$. Table~\ref{tab1} lists $N^{-1}\ln{\lambda(N,n)}$ and $W(N,N_e)$ at $n = 1$ and $1/2$. 
The dimension of the basis involves counting and holds for arbitrary interactions. The size dependence is still appreciable at $N \sim 100$.

\begin{table}[t]
  \caption{Number of states $W(N,N_e)$, Eq.~\ref{eq:finitebasis}, and $A(N,n) = N^{-1}\ln{\lambda(N,n)}$ for $N$ sites and $N_e = N$ or $N/2$.}
  \label{tab1}
  \begin{tabular}{lllll}
    \hline
    \hline
$N$ & $ W(N,N)$ & $A(N,1)$ & $W(N,N/2)$ & $A(N,1/2)$ \\
    \hline
8  &         12870  &  0.203 &      1820 &  0.186  \\
10 &        184756  &  0.174 &      15504 & 0.160 \\
12 &         2.70E+06 &      0.152 & 134596 &        0.141  \\
16 &         6.01E+08 &      0.123 & 1.05E+07 &      0.114   \\
20 &         1.38E+11 &      0.104 & 8.48E+08 &      0.0968   \\
28 &        7.65E+15 &      0.0801 &        5.81E+12 &      0.0750   \\
36 &        4.43E+20 &      0.0658 &        4.14E+16 &      0.0618   \\
48 &        6.44E+27 &      0.0523 &        2.61E+22 &      0.0493    \\
64 &        2.40E+37 &      0.0415 &        1.48E+30 &      0.0392   \\
96 &        3.61E+56 &      0.0297 &        5.15E+45 &      0.0283    \\
\hline
\end{tabular}
\end{table}

The canonical partition function $Q(T,N,N_e)$ is the Boltzmann weighted sum over the spectrum $\{E(N,N_e)\}$ 
of models with $N$ sites and $N_e$ electrons. ED returns the full spectrum of correlated states of small systems. 
The scaled partition function with $\beta = 1/T$ is
\begin{equation}
	{Q}_{\lambda} \left (T,N, {N}_{e} \right ) =\lambda(N,n) \sum_{j} \exp{-\beta {E}_{j} (N, {N}_{e})}. 
\label{eq:scaled_partition}
\end{equation}
Scaling ensures that the numerical value of  $N^{-1}$ $\ln{Q_\lambda(T,N,N_e)}$ in the high $T$ limit is $\ln{g(n)}$ in Eq.~\ref{eq:scale_factor}. 
The details of $\{E(N,N_e)\}$ do not matter when $T$ is large compared to energy differences and the sum in Eq.~\ref{eq:scaled_partition} becomes 
an integral over excitation energies. Of course, finite-size gaps inevitably become important at low $T$ when, however, only low-energy 
excitations contribute to the partition function. 

We compute the entropy per site and use $S_\lambda(T,N,N_e)$ for the scaled entropy,
\begin{equation}
	{S}_{\lambda} (T,N, {N}_{e}) =S(T,N, {N}_{e}) + {N}^{-1} \ln{\lambda(N,n)}.    
\label{eq:scaled_entropy}
\end{equation}
As shown below, $S_\lambda(T,N,N_e)$ greatly facilitates finding the thermodynamic limit $S(T,n)$ even though the energy 
spectrum of finite systems is retained. First, we address the difference between the band and ED/DMRG $\chi(T)$ in the lower panel of Fig.~\ref{fig1}.

The energy spectrum of Eq.~\ref{eq:EHM_ham} in a static magnetic field $h$ has Zeeman components $E_j(N,N_e) - mhS_j$ with $m$ 
ranging from $-S_j$ to $S_j$ in states with spin $S_j$. The susceptibility is given by derivatives of $\ln{Q(T,N,N_e,h)}$ evaluated at $h = 0$. 
Nevertheless, $\chi(T,N,N_e)$ depends on system size. The high $T$ limit of $\chi(T,N,N_e)$ is a Curie law, $1/4T$ in reduced units, 
for a density $\rho_1(N,N_e)$ of $S = 1/2$ sites. The size dependence of $\rho_1(N,N)$ follows from Eq.~\ref{eq:finitebasis},
\begin{equation}
	{\rho}_{1}(N,N) = \sum_{q=0}^{{N} / {2}} {\left(1-\frac {2q} {N} \right) \frac{W(q,N,N)} {W \left(N,N \right)}}.
\label{eq:density_halffill}
\end{equation}
We have $\rho_1 = 0.53333$ and $0.52631$ at $N = 8$ and $10$, larger than the thermodynamic limit $\rho_1(1) = 0.50$. Hence $\chi(T,14)$ in Fig.~\ref{fig1} 
is larger than $\chi(T)$ at high $T$ by a known amount that can be verified quantitatively and can explicitly be shown to depend on 
system size but not on interactions. Deviations at $T \sim 1$ occur in larger systems with, for example,  $\rho_1 = 0.51064$ and $0.50526$ at $N = 24$ 
and $48$, respectively. It follows that $\chi(T,N) \to \chi(T)$ with increasing $N$, although numerical results are then 
limited to very low $T$. We approximate the thermodynamic limit of $\chi(T)$ to an accuracy shown in Fig.~\ref{fig1} for noninteracting 
fermions with ED to $N = 14$. The high $T$ limit of finite spin chains, by contrast, returns the thermodynamic limit 
because $N^{-1}\ln{W(N)} = (2S + 1)$ is size independent without scaling.
\begin{figure}
\includegraphics[width=\columnwidth]{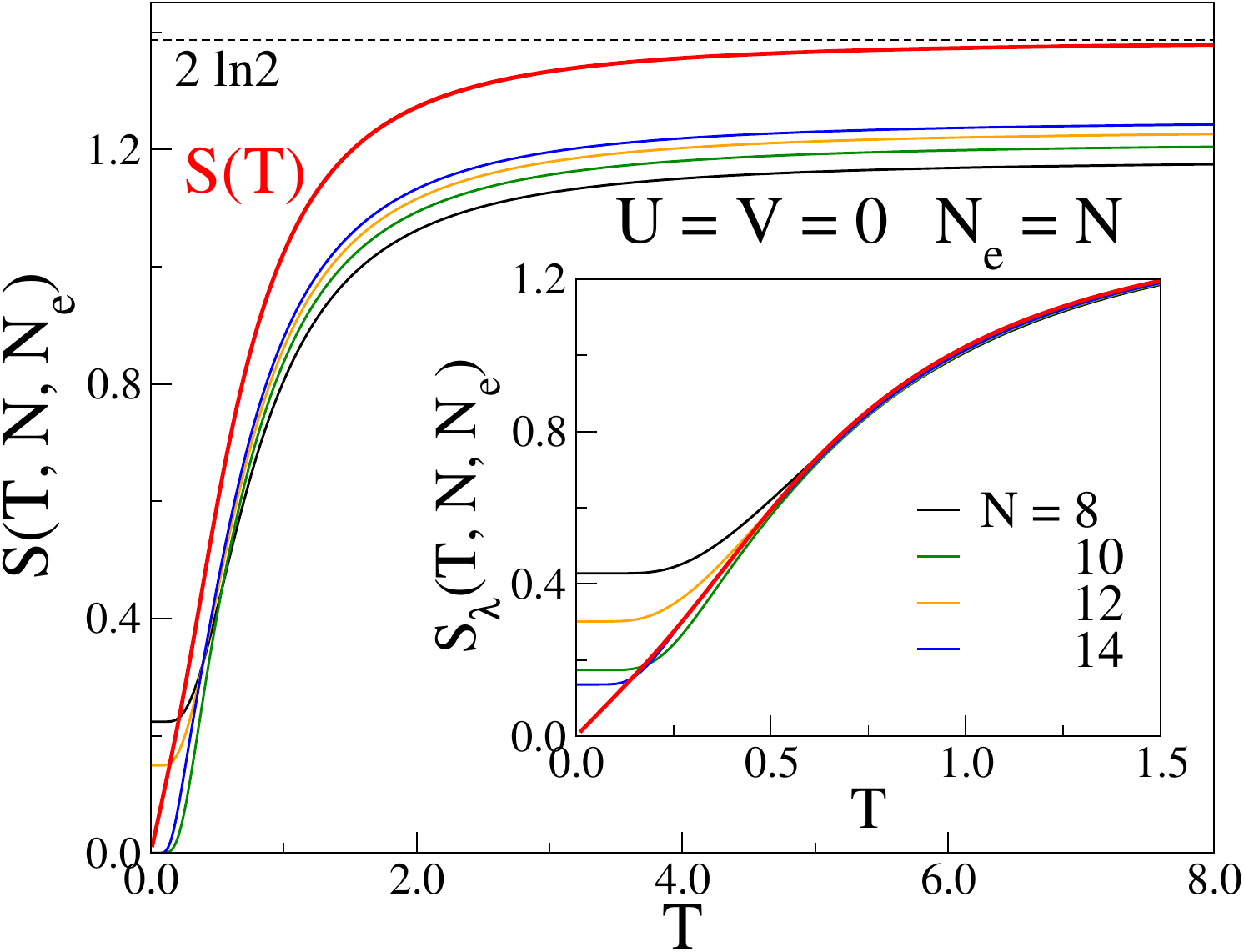}
\caption{\label{fig2}
Exact entropy per site $S(T,N,N_e)$ of systems with $N = N_e \le 14$ and $U = V = 0$ in Eq.~\ref{eq:EHM_ham}; the band limit is $S(T)$. 
	Inset: Scaled $S_\lambda(T,N,N_e)$ and $S(T)$ up to $T = 1.5$. The high $T$ limit of $S(T)$ or $S_\lambda(T,N,N_e)$ is $2ln2$.}
\end{figure}

Fig.~\ref{fig2} shows exact entropy densities of half-filled noninteracting systems. The band limit (red curve) is 
initially linear in $T$ and reaches $2\ln{2}$ at high $T$. The $N = 8$ and $12$ curves start at $S(0,4p) = (4p)^{-1}\ln{6}$ and cross 
at finite $T$. The $N = 10$ and $14$ curves have finite size gaps and $S(T,14)$ is larger than $S(T,10)$ at all $T$. The inset shows the 
scaled entropy $S_\lambda(T,N,N_e)$ at the same system sizes. The thermodynamic limit $S(T)$ is reached at $T > 0.6$ with $N = 4p$ 
converging from above and $N = 4p + 2$ crossing $S(T)$ at low $T$. An upward shift of 
$S(T,N,N)$ by $N^{-1}\ln{\lambda(N,1)}$ 
returns $S(T)$ for $T > 0.6$. In this case, $S_\lambda(T,14,14)$  is an excellent approximation to $S(T)$ for $T > 0.15$.

Finite models have $S_\lambda(0,N,N_e) > 0$ while gapless models with a nondegenerate ground state have linear $S(T)$ 
close to $T = 0$. It follows that the $S_\lambda(T,N,N_e)$ and $S(T)$ curves cross at low $T$ in gapless models with $N_e = 4p + 2$. 
Entropy conservation and convergence to $S(T)$ at $T > T(N,N_e)$ then ensure convergence from below to the thermodynamic 
limit with increasing system size. Aside from the crossing region at low $T$, models with a nondegenerate ground states and $N_e = 4p + 2$ satisfy
\begin{equation}
{S}_{\lambda} \left (T, {N}_{e} \right ) \leq  S \left (T,n \right ).
\label{eq:entropy_conv_below}
\end{equation}
Let us suppose that DMRG returns the energy spectrum $E(N,N_e) \leq E_C(N,N_e)$ up to a cutoff that leads to a truncated $S_C(T,N,N_e)$. Since Boltzmann factors and excitation 
energies are non-negative, truncation cannot increase the entropy. The crossings shown in the inset for noninteracting fermions provide a general 
approach to correlated systems: Truncated $S_\lambda(T,N_e)$ with $N_e = 4p + 2$ converge from below to $S_\lambda(T,N_e)$ on 
increasing $E_C(N)$ and $S_\lambda(T,N_e)$ converges from below to $S(T,n)$ on increasing $N$. In principle, $S(T,n)$ can 
be obtained by first increasing $E_C(N)$ and then $N$.

\subsection{\label{sec2b} DMRG}



\begin{figure}[t]
	\includegraphics[width=\columnwidth]{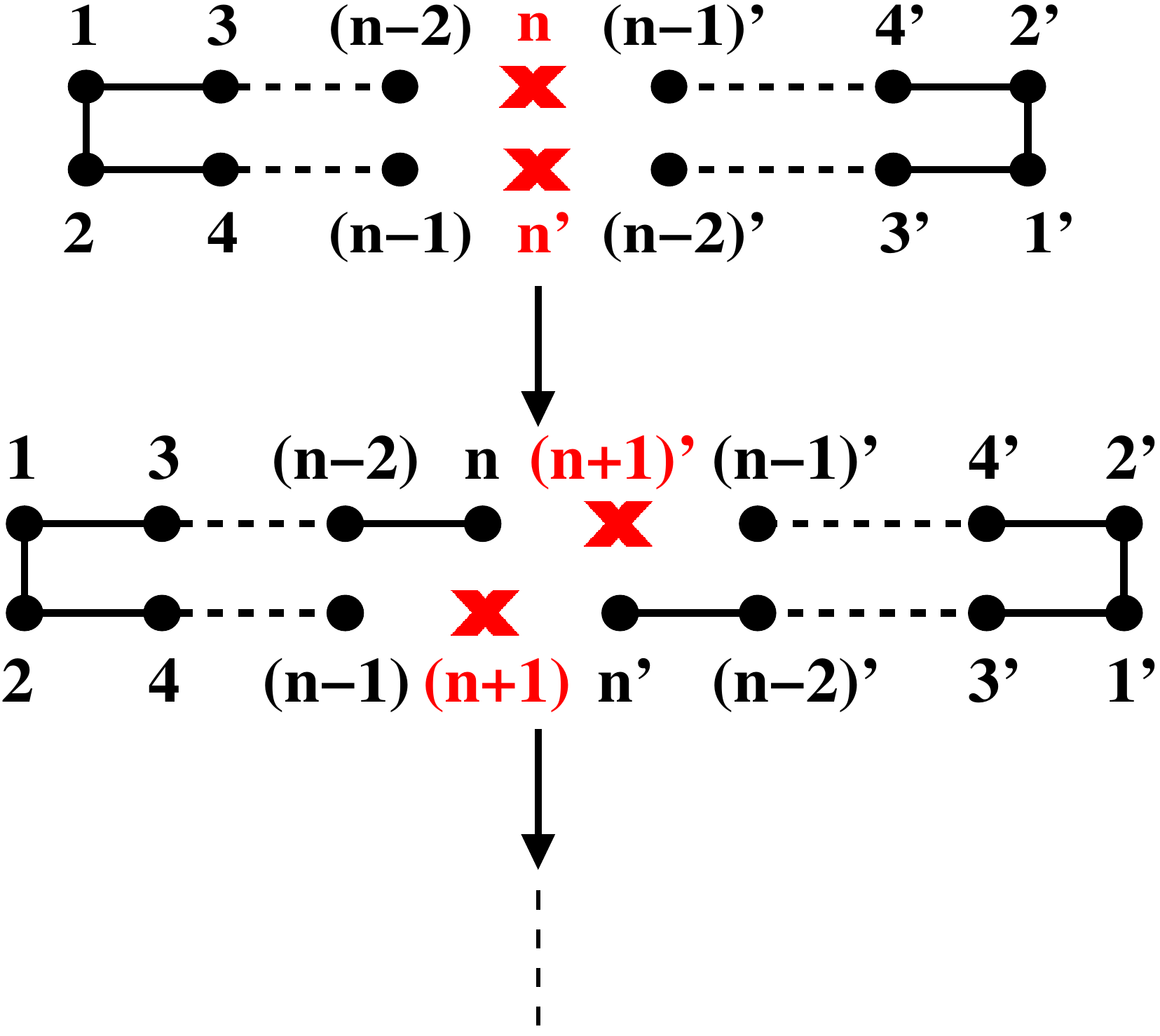}
\caption{\label{fig3}
Infinite DMRG algorithm for a system with periodic boundary conditions. Left and right block sites are numbered as unprimed and 
	primed integers, respectively. Sites of left and right blocks are shown as black circles and new sites added at DMRG steps 
	are represented by red crosses.}
\end{figure}

The models in Eq.~\ref{eq:EHM_ham} conserve $S$ and $S^Z \leq N_e/2$. DMRG calculations are performed in sectors with fixed $S^Z$ using an 
algorithm for periodic boundary conditions similar to Ref.~\citenum{mkumar2009}. The superblock consists of two new sites, a left new and 
right new site in addition to the left and right blocks. For improved implementation of periodic boundary conditions, we modified the addition of 
new sites in the infinite DMRG algorithm. New sites are now added alternately at each end of the left and right block as shown in the Fig.~\ref{fig3}. The left and right 
block are increased by one site at each step of infinite DMRG until the desired system size $N$ is reached. We chose $m = 500$ eigenvectors of the density matrix of 
the system block that corresponds to the highest eigenvalues after testing $m = 300$ and $400$. The dimension of the superblock, the Hamiltonian matrix, is $m^2 4^2$,  
and the computational cost to diagonalize the Hamiltonian goes as O($m^3$). To optimize the efficiency our calculation we use $m \sim 500$ and  $5 - 10$ sweeps 
of finite DMRG for all calculations. 

We seek the low energy excitations $E_p(N)$ at system size $N$ instead of the ground state. To improve the accuracy of the low lying spectrum, we construct 
the system block density matrices $\rho_l(N)$ with  $l$ targeted energy levels of the superblock at system size $N$ and define an effective density $\rho^\prime(\beta^\prime,l)$
\begin{equation}
\rho^\prime(\beta^\prime,l,N)=\sum_{p=1}^l \frac{\rho_p (N)  \exp{[-\beta^\prime E_p(N)]}} {Q_l(T,N)}.
\label{eq:dm_renorm}
\end{equation}
The $l = 1$ case is $\rho^\prime(\beta,1) = \rho_1$ when the ground state is sought. Contributions for $l > 1$ are governed by $\beta^\prime$, 
an effective inverse $T$. We set $\beta^\prime = 10$ (in units of $1/t$) since $T \sim 0.1$ is the range of interest. Variations of $\beta^\prime$ 
by $10\%$ to $20\%$ hardly change the accuracy of the spectrum. The effective density matrix becomes important when the lowest excitations are closely 
spaced and projections of all the degenerate states of the superblock have equal contribution in constructing the density matrix of the system block. 

The system block Hamiltonian and all operators are renormalized by $\rho^\prime(\beta,l,N)$ to obtain the energy spectrum of the model Hamiltonian at 
system size $N$. We perform two calculations. We first take $l = 5$ or $10$ in order to obtain the lowest excitations very accurately. The second 
calculation has $l > 100$ (most have $l = 200$). The entire spectrum is red shifted by approximately a constant amount because the 
density matrix now has projections from many excited states. Accordingly, we shift back the spectrum by a constant and use the first calculation 
for the lowest excitations. 

The absolute ground state of the models studied is in the $S^Z = 0$ sector, which contains a Zeeman component of all states with $S > 0$. 
The sectors $S^Z > 0$ are doubly degenerate and $l = 200$ contains states at higher energy than $l = 200$, $S^Z = 0$ or $S^Z = 1$. As shown in 
Section~\ref{sec2c}, fewer states suffice for converged thermodynamics at low $T$.

We have already mentioned the single-particle states $\varepsilon(k) = -2\cos{k}$, $ -\pi < k \le \pi$, of Eq.~\ref{eq:EHM_ham} with $U = V = 0$ at system size $N$. 
The ground-state degeneracy in the $S^Z = 0$ sector is 4 or 1 for $N_e = 4p$ or $4p + 2$, respectively. The exact excitation spectrum for $S^Z = 0$ and its degeneracy 
are easily obtained for a single electron-hole pair or for two e-h pairs. We compare exact and DMRG excitations in the $S^Z = 0$ sector in the 
Appendix for six representative half-filled and quarter-filled systems. The exact excitations $E_{p-p^\prime}$ are degenerate from states $p$ to $p^\prime$. 
The DMRG accuracy decreases from a percent or so at $p$ to about $10\%$ to $20\%$ at $p^\prime$, as seen both for different $E_{p-p^\prime}$ at fixed $N$ and 
same $E_{p-p^\prime}$  at different $N$. We also notice that the accuracy decreases from $p$ to $p^\prime$ for highly degenerate excitations.

As in standard DMRG, accuracy and computational effort increase with $m$. We found $m = 500$ to be clearly superior to 300 or 400 and almost comparable to 
$m = 600$ in some test calculations. The choice of $l$ or $\beta^\prime$ in Eq.~\ref{eq:dm_renorm} for targeting correlated spectra at system size $N$ 
is less tested. Thermodynamic comparisons with band theory or with exacts results for correlated systems are more appropriate in our opinion than excitation 
spectra, except for the spin or charge gaps that are accessible to standard DMRG.

\subsection{\label{sec2c} Energy cutoff}

We seek a cutoff $E_C(N,N_e)$ at system size $N$ and retain all states $R(N,N_e)$ with
$E_p(N,N_e) \le E_C(N,N_e)$. The cutoff must be high enough to suppress finite-size effects and yet low enough for the truncated spectrum to return converged thermodynamics in a narrow range $T(N)$ at each system size.

\begin{figure}
        \includegraphics[width=\columnwidth]{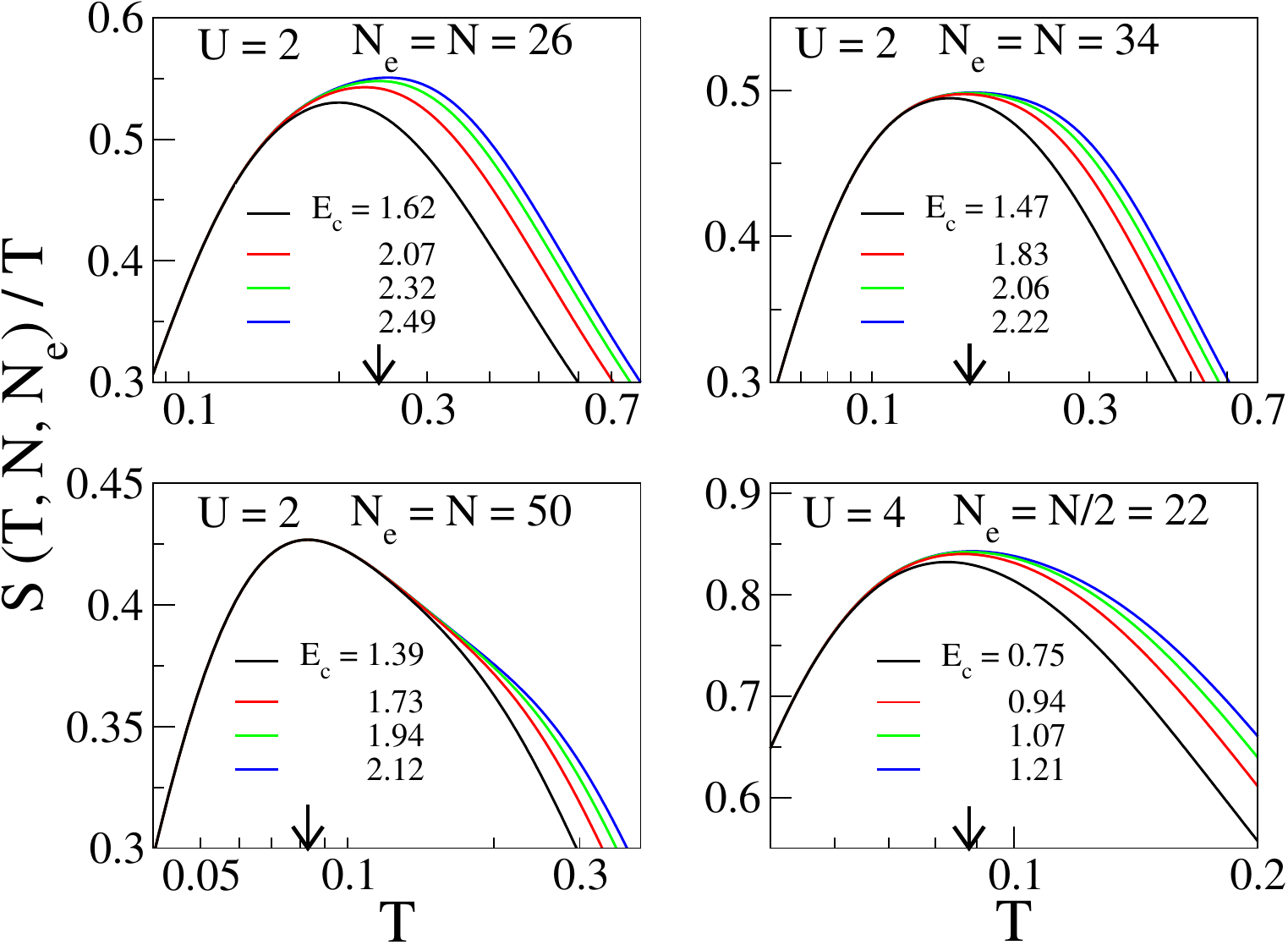}
\caption{\label{fig4}
Dependence of truncated $S(T,N,N_e)/T$ in the energy cutoff $E_C$ in systems with $N$ sites, $N_e$ electrons, $V = 0$ and $U$ in Eq.~\ref{eq:EHM_ham}. 
	The arrows mark $T(N)$, the maximum of $S(T,N,N_e)/T$ for $E_C$ corresponding 
the green curves.}
\end{figure}

We typically increase $E_C(N,N_e)$ to include $200-300$ states in the $S^Z = 0$ sector that contains the ground state. $R(N,N_e)$ is three or four times larger due to contributions from sectors with Zeeman degeneracy $2S + 1$. The truncated entropy $S_C(T,N,N_e)$ 
has $E_j(N,N_e)\leq  E_C(N,N_e)$ in Eq. \ref{eq:scaled_partition}. The cutoff
criterion~\cite{sudip19} is to converge or almost converge the maximum of $S_C(T,N)/T$. Fig.~\ref{fig4} shows the dependence of $S_C(T,N,N_e)/T$ on $E_C(N,N_e)$ 
for representative cases, three panels with $N_e = N$ and one with $N_e = N/2$. As expected, convergence to $S_C(T,N,N_e)$ is excellent 
at low $T$. Once reached, the maximum $T(N)$ marked by arrows becomes insensitive to
additional increase of the cutoff.

The cutoffs used in most of the following calculations are the green curves in Fig.~\ref{fig4}, the second highest $E_C$. Table~\ref{tab2} 
lists $E_C(N,N_e)$, $T(N)$ and $R(N,N_e)$. Higher $E_C$ is an option but is time 
intensive. As also found for noninteracting fermions in the Appendix, $T(N)$ is close 
to the finite size gap $\Delta(N)$ at system size $N$. Convergence to the thermodynamic 
limit requires, quite reasonably, $T(N) \sim \Delta(N)$. The DMRG excitations are the
most accurate in this range. 
With $E_C/T(N) \sim 10$ at $U = 2$ in Table~\ref{tab2}, 
the Boltzmann factors of neglected states are less than $5 \times 10^{-5}$ and are also small for less accurate excitations close to the cutoff.

 \begin{table}[b]
	 \centering
 \caption{Number of states $R(N,N_e)$ in systems with $N_e = N$ electrons, $U = 2$ and $V = 0$ up to the cutoff $E_C(N)$. $\Delta(N)$ is the finite
size gap; $T(N)$ is discussed in the text.}
  \label{tab2}
  \begin{tabular}{lllll}
    \hline
    \hline
$N$  & $ T(N)$  & $E_C(N)$ &  $R(N,N_e)$  & $\Delta(N)$  \\ 
    \hline
        14 &    0.490 & 3.03 &  927 & 0.537   \\
        18 &    0.320 & 2.7 &   591 & 0.416 \\
        22 &    0.271 & 2.42 &  481 & 0.342  \\
        26 &    0.240 & 2.32 &  402 & 0.292  \\
        30 &    0.195 & 2.1 &   501 & 0.245  \\
        34 &    0.164 & 2.06 &  532 & 0.217  \\
        50 &    0.083 & 1.94 &  611 & 0.143  \\
        66 &    0.056 & 1.92 &  713 & 0.106  \\
    \hline
  \end{tabular}
\end{table}

We obtain the thermodynamic limit $S(T,n)$ by interpolation of DMRG results. $S(T)$ is initially linear in $T$ in gapless models. 
Its thermodynamic limit at high $T$ is given by $S_\lambda(T,N)$ in Fig.~\ref{fig2}, inset. DMRG yields $S_\lambda(T,N)$ up to $T(N)$ 
when the spectrum is truncated at $E_C$. At intermediate $T$, we interpolate $S_\lambda(T,N)$  between higher $T$ and lower $T$. 
Convergence to $S(T,n)$ is from below for $N = 4p + 2$. It follows that $S(T,n)$ is given by $S_\lambda(T,N,N_e)$ with progressively larger $N$ at lower $T$.

\begin{figure}[t]
        \includegraphics[width=\columnwidth]{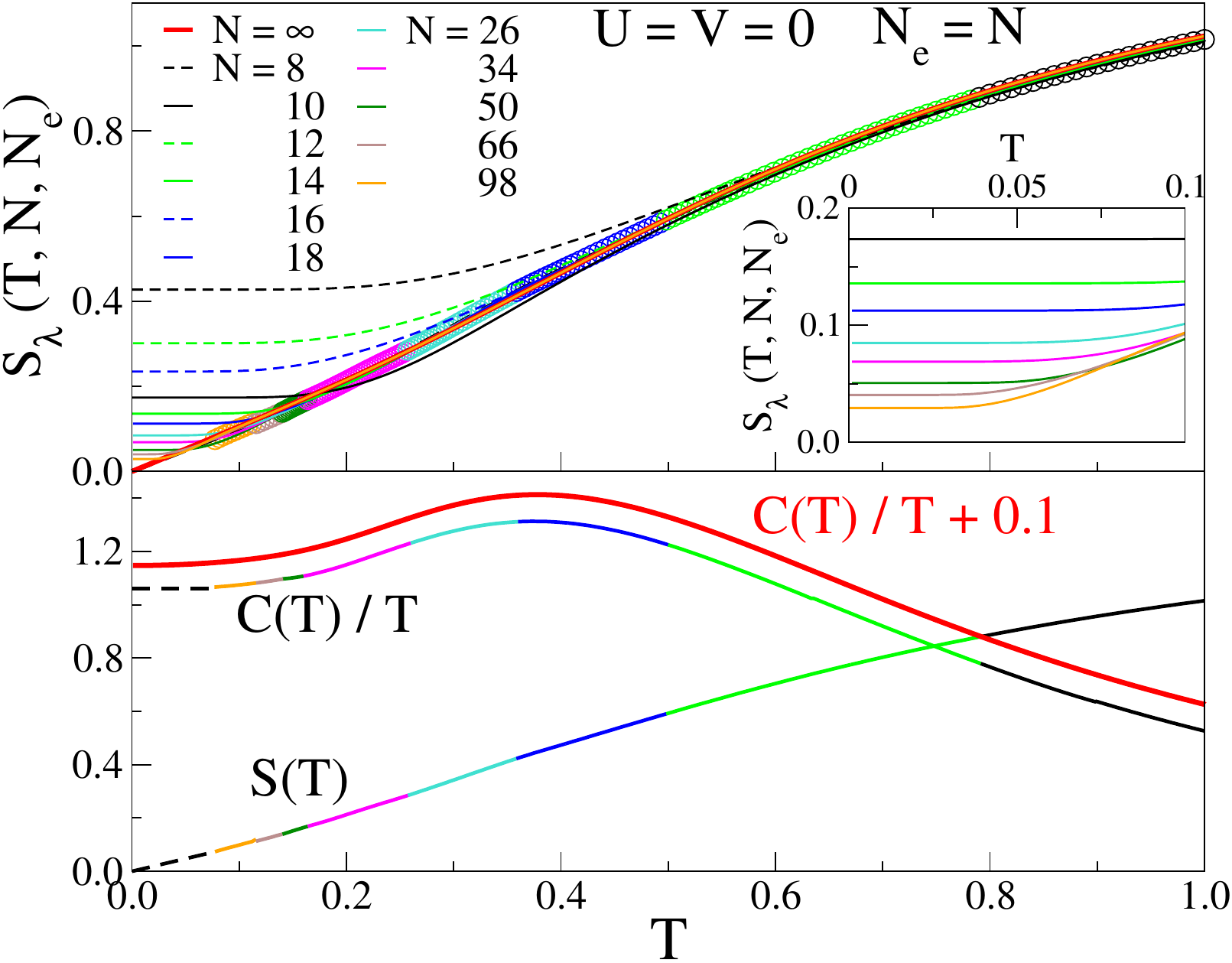}
\caption{\label{fig5}
Upper panel: Scaled entropy $S_\lambda(T,N,N_e)$ per site at $T < 1$ for $N = N_e$ and $U = V = 0$ in Eq.~\ref{eq:EHM_ham}. The full 
	red line is the band limit. The color coding indicates the system size that approximates the thermodynamic limit. The 
	inset zooms in on $T \leq 0.1$. Lower panel: The calculated $S(T)$ and its derivative $S^\prime(T) = C(T)/T$; 
	the band limit $C(T)/T$ is offset by 0.10 for clarity.}
\end{figure}

The upper panel of Fig.~\ref{fig5} shows $S_\lambda(T,N,N_e)$ for $U = V = 0$ and $N_e = N$ up to $T = 1$, beyond which ED returns the 
thermodynamic limit. The solid red line is the band limit. The inset zooms in on $T \leq  0.1$. The $N \leq  14$ curves are ED and 
converge from above (below) for $N = 4p$ ($4p + 2$). The $N = 4p + 2$ systems are lower 
bounds. The color-coded points indicate the system sizes that give the calculated $S(T)$ shown in the lower panel. Agreement with the 
band limit is almost quantitative; the largest difference in Fig.~\ref{fig5} is 0.006. The entropy derivative is $S^\prime(T) = C(T)/T$ 
where $C(T)$ is the specific heat per site. The calculated $S^\prime(T)$ is compared in the lower panel to the band limit, offset by 0.10.

The entropy analysis in Fig.~\ref{fig5} came as a surprise. With good reason, 
deviations from the thermodynamic limit have been associated with finite size gaps. 
That remains the case for spin chains, which have indeed been modeled the most. However, 
the size dependence of the many-fermion basis is just as important as gaps for the 
entropy or free energy of fermionic models. Scaling the partition function by
$\lambda(N,1)$ to ensure the proper high $T$ limit shifts the $S(T,N,N)$
curves to $S(T)$ at progressively lower $T$ with increasing $N$.

The ED/DMRG procedure illustrated above for $U = 0$ and $N_e = N$ is accurate down to at least $T \sim 0.03$ for both $\chi(T)$ 
in Fig.~\ref{fig1} and for $S(T)$ and $S^\prime(T)$ in Fig.~\ref{fig5}. The method tends to perform better in models with less 
dramatic $N_e = 4p$, $4p + 2$ variations. Larger systems are computationally more demanding. The actual limit~\cite{sudip19} is set by 
the accuracy of the dense spectrum of large systems, which is highly model dependent. 

\section{\label{sec3} Representative results}

We apply the ED/DMRG method to fermionic models such as Eq.~\ref{eq:EHM_ham} and seek accurate low $T$ thermodynamics. The size dependence of the entropy $S(T,N)$ per site
generates $T(N)$ in Table~\ref{tab2} at which the thermodynamic limit is reached. The
ED/DMRG results in the lower panel of Fig.~\ref{fig1} are based on the crossing or
merging points $T_N$ of $N = 4p$, $4p + 2$ susceptibilities. The $96/98$ curves cross at $T_{98} = 0.118$, slightly below the truncated $S(T,98)/T$ maximum at $T(98) = 0.14$, as
required to be in the thermodynamic limit for $T > T(N)$. Gapless systems have finite 
$\chi(T)$ and $S^\prime(T)$ at $T = 0$ 
while gapped systems have $\chi(0) = S^\prime(0) = 0$ and $E_{ST} > 0$ to the lowest triplet. Since $E_{ST}(N)$ decreases with N in finite systems, it follows 
that $\chi(T,N_e)$ converges to $\chi(T,n)$ from 
below for $N_e = 4p + 2$,
\begin{equation}
\chi \left (T, {N}_{e} \right ) \leq \chi \left (T,n \right )  .
\label{eq:suscep_conv_below}
\end{equation}
The calculated $\chi(T)$ based on merging at $T_N$ 
and Eq.~\ref{eq:suscep_conv_below} are closely similar, with maximum deviation of $3 \times 10^{-4}$ at $T = 0.13$ or $0.27$. 
We use merging points, $4p + 2$ convergence and $T(N)$ in the following. They
exploit the size dependence and return consistent thermodynamic limits.

\subsection{\label{sec3a}Hubbard model, $U < 4$}
\begin{figure}
        \includegraphics[width=\columnwidth]{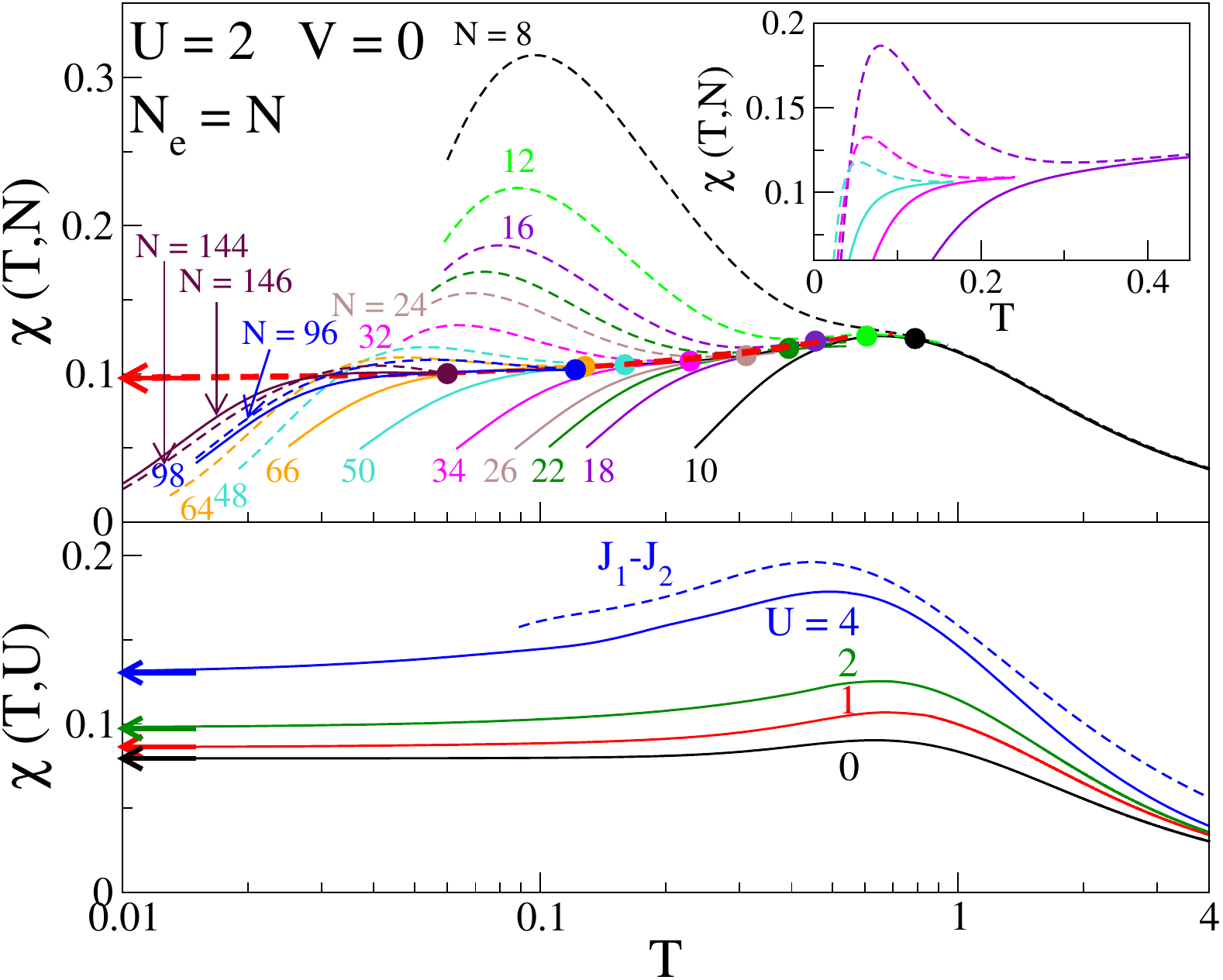}
\caption{\label{fig6}
Upper panel: Same as Fig.~\ref{fig1} for $U = 2$ instead of $U = 0$; $\chi(T,N)$ for $N = 4p$, $4p + 2$ 
	merge at the indicate points that give $\chi(T,U)$ in the thermodynamic limit. The inset shows the merging of $N = 16/18$, 
	32/34 and 48/50. Lower panel: $\chi(T,U)$ at $U = 0$, 1, 2 and 4. The arrows in both panels are exact $\chi(0,U)$ from Ref.~\citenum{takahashi70}. 
	The dashed line is ED for $N = 24$, $J_1 = 3/4$ and $J_2 = 1/16$ in Eq.~\ref{eq:j1j2}.} 
\end{figure}

We start with half-filled Hubbard models with $U \leq 4$, less than or equal to the bandwidth, and $V = 0$ in Eq.~\ref{eq:EHM_ham}. 
Fig.~\ref{fig6}, upper panel, shows $\chi(T,N)$ at $U = 2$ using ED up to $N = 12$ and DMRG for larger systems. The notation follows 
Fig.~\ref{fig1}. The bold dashed line is the thermodynamic limit $\chi(T,U)$ inferred from the indicated merging points.  
Since the ground state is a singlet for $U > 0$, small $N = 4p$ systems merge from above as in the band limit. The $\chi(T,4p)$ maxima 
due to finite size decrease and almost disappear by $N = 144$. The $N = 4p + 2$ curves that converge from below yield accurate $\chi(T)$ 
to $T \sim 0.02$ or less for $N = 146$, well below the $144/146$ merging point. 

The lower panel presents calculated $\chi(T,U)$ at $U = 0$, 1, 2 and 4. The arrows mark the exact $\chi(0,U)$ obtained analytically by 
Takahashi~\cite{takahashi70}. All $\chi(T,N)$ increase with $N$ at $U = 4$ and converge from below to $\chi(T)$. 
Finite $U > 0$ increases $\chi(T,U)$ at low $T$ since the ground and low-energy states have reduced contribution from doubly 
and unoccupied sites, still appreciable at $U = 4$, whereas spin chains have exclusively singly occupied sites. 
The exchanges in Eq.~\ref{eq:fermion_j1j2} at $U = 4$ are $J_1 = 3/4$ and $J_2 = 1/16$. ED for $N = 24$ gives the spin $\chi(T)$ shown as a dashed 
line in the lower panel for $T > 0.09$ where the thermodynamic limit holds. Sites with $n_r = 2$ or 0 reduce $\chi(T,U)$ at low $T$. 
The reduction becomes negligible by $U = 8$ where the HAF with $J_1 = 1/2$ and $J_2 = 0$ is almost quantitative. Exact HAF results have long been used to benchmark numerical methods.
\begin{figure}
        \includegraphics[width=\columnwidth]{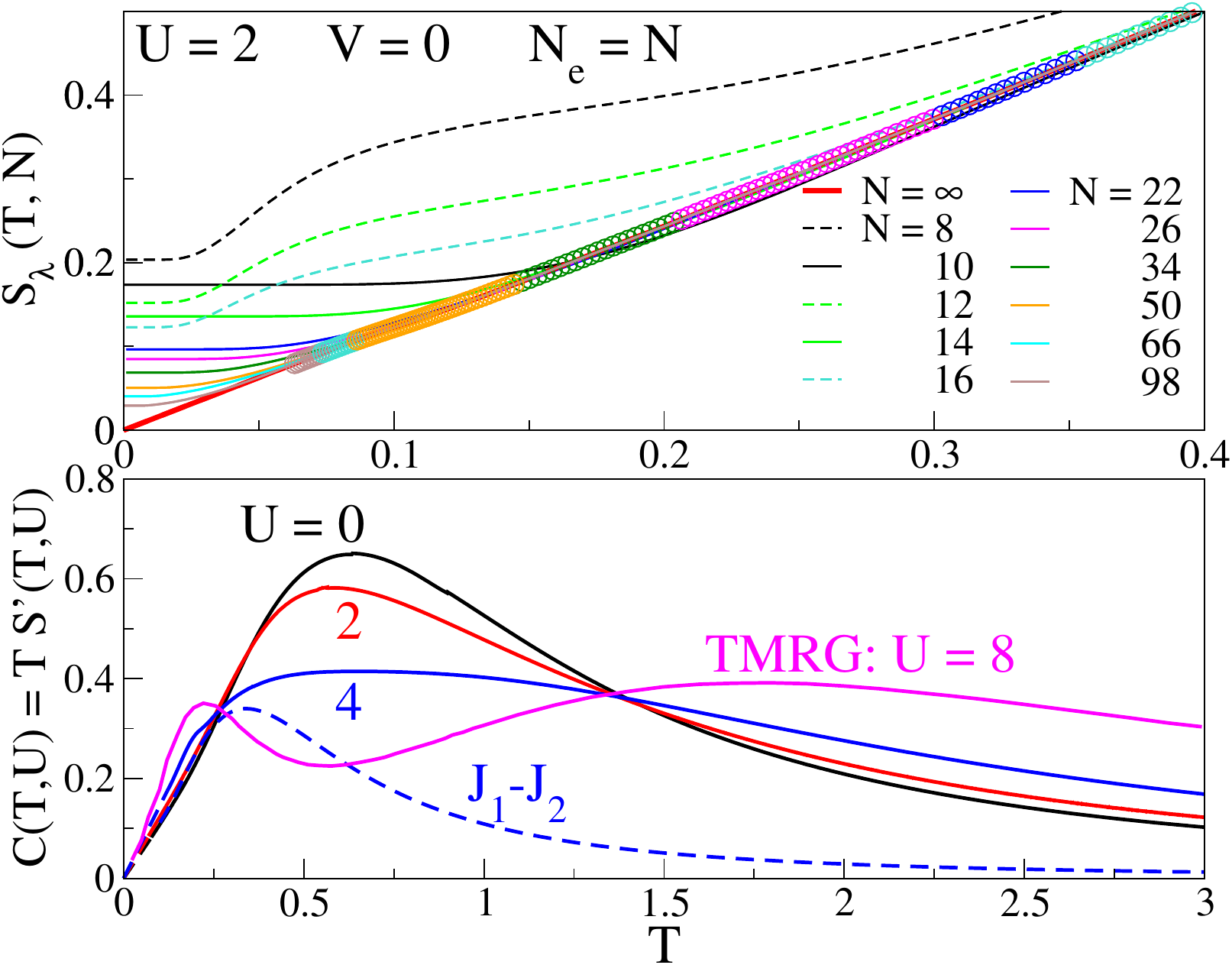}
\caption{\label{fig7}
Upper panel: Scaled entropy $S_\lambda(T,N)$ per site at $T \leq 0.40$ for $N = N_e$ and $U = 2$, $V = 0$ in Eq.~\ref{eq:EHM_ham}. 
	The full red line is the thermodynamic limit. The color coding indicates the system size that approximates the thermodynamic limit. 
	Lower panel: The calculated $C(T,U) = TS^\prime(T,U)$ at $U = 0$, 2 and 4. The dashed $J_1-J_2$ line is ED for 24 spins 
	with exchanges at $U = 4$ in Eq.~\ref{eq:fermion_j1j2}. The TMRG curves at $U = 8$ is from Ref.~\citenum{sirker2007}.}
\end{figure}

Fig.~\ref{fig7}, upper panel, shows $S_\lambda(T,N)$ at $U = 2$ and $T \leq 0.4$ for the indicated system sizes.
ED at $N = 10$ and 12 for the scaled entropies converge to $S(T,U)$ for $T > 0.4$ as shown in Fig.~\ref{fig2}, inset, for $U = 0$. 
Small $N = 4p$ gaps initially increase the entropy. As in Fig.~\ref{fig5}, the thermodynamic limit $S(T,U)$ is color coded 
according to the system size that contributes at $T$. Systems up to $N = 98$ determine $S(T,U)$ directly for $T > 0.06$ 
and by extrapolation at lower $T$.

The lower panel of Fig.~\ref{fig7} compares the calculated specific heat $C(T,U) = TS^\prime(T,U)$ at $U = 0$, 2 and 4. 
The dashed line for $T > 0.09$ is the spin chain, Eq.~\ref{eq:j1j2}, with $J_1 = 3/4$, $J_2 = 1/16$ at $U = 4$. 
Its $C(T)$ maximum is exclusively due to spins. The $C(T,U)$ maxima for $U = 0$ and 2 have contributions from spin, charge 
and combined excitations. Increasing $U$ shifts charge excitations to higher energy, thereby reducing the maximum. 
Combined spin and charge excitations lead to the broad $C(T,4)$ curve. We include the TMRG curve at $U = 8$ from Ref.~\citenum{sirker2007} 
that illustrates spin-charge separation. As discussed in Section~\ref{sec3b}, ED for $N = 8$ or 10 at $U = 8$ also returns two $C(T)$ peaks. The low $T$ thermodynamics at $U \geq 4$ are more reliably based on spin chains, Eq.~\ref{eq:j1j2}, even though it only becomes truly quantitative in the large $U$ limit.

\subsection{\label{sec3b}Spin-charge separation, $U > 4$}

The spin-1/2 chains in Eq.~\ref{eq:j1j2} describe the thermodynamics of Eq.~\ref{eq:EHM_ham} in half-filled systems with $U > 4$ or $U - V > 4$ 
when interactions exceed the bandwidth $4t$. The atomic limit $U \to \infty$ has noninteracting spins with Curie susceptibility 
of $\chi_{Curie}(T) = 1/4T$ while the band limit is $1/8T$ since half of the sites are singly occupied at high $T$. 

\begin{figure}[]
        \includegraphics[width=\columnwidth]{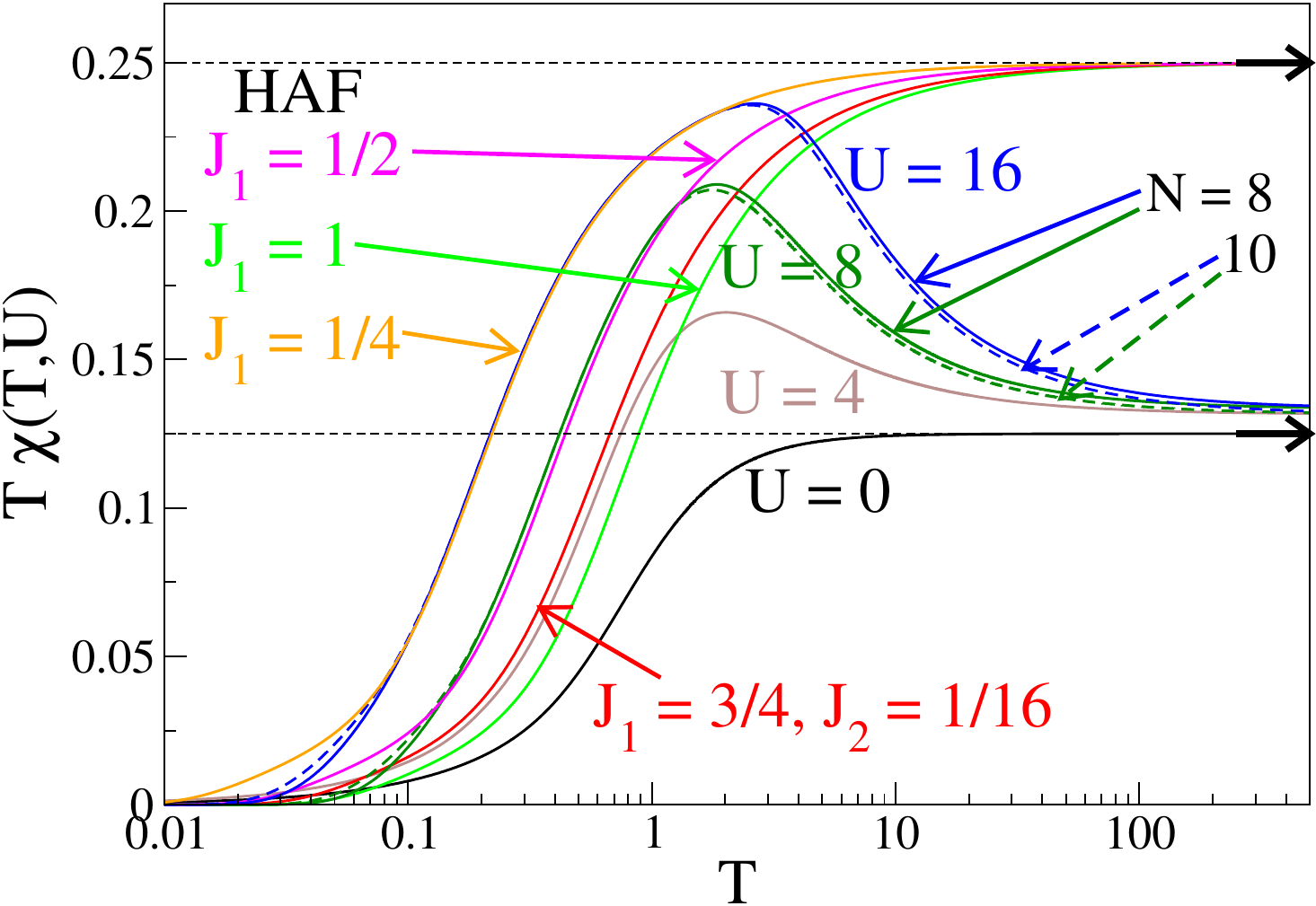}
\caption{\label{fig8}
	Calculated $T\chi(T,U)$ at $U = 0$ (band limit) and $4$ in the thermodynamic limit and at $U = 8$ and $16$ 
	for $N = 8$ and 10. The spin chains with $T\chi(T) \to 1/4$ at high $T$ are ED for 24 spins: HAFs with $J_1 = 1$, 
	1/2 and 1/4 and the $J_1-J_2$ model with $J_1 = 3/4$, $J_2 = 1/16$.}
\end{figure}

The $U = 0$ line in Fig.~\ref{fig8} is $T\chi(T)$, the band limit in Eq.~\ref{eq:pauli_paramag}, that increases monotonically to 1/8. 
The $U = 4$ line is the thermodynamic limit $T\chi(T,U)$ from Fig.~\ref{fig6}, lower panel. The maximum at $T \sim 1.8$ is due to spin 
excitations. The full and dashed lines at $U = 8$ and 16 are $T\chi(T,U,N)$ at system sizes $N = 8$ and 10. The maxima become more pronounced 
and shift to lower energy. Finite size effects are evident at both low and high $T$ where we have $T\chi(T) > 1/8$ because $\rho_1(N) > 1/2$ 
as discussed in connection with Fig.~\ref{fig1}. The ED/DMRG method is not applicable to systems with spin-charge separation. 
The low-energy states accessible by DMRG limit convergence to $T \leq T(N) < 1/2$ in Table~\ref{tab2}.

We turn instead to spin chains and ED for $N = 24$ spins in Eq.~\ref{eq:j1j2} to compute $T\chi(T)$. The thermodynamic limit 
holds for $T > 0.1$ before any DMRG calculation. The exchanges are $J_1 = 3/4$ and $J_2 = 1/16$ at $U = 4$. To lowest order in $(t/U)^2$, 
we have $J_2 = 0$ and an HAF with $J_1 = 1$. The impact of $J_2$ on $T\chi(T)$ is evident in Fig.~\ref{fig8}. The HAF substantially 
underestimates the $U = 4$ susceptibility at low $T$ while the $J_1-J_2$ model slightly overestimates it as has already been noted. The HAF 
has $J_1 = 1/2$ and 1/4 at $U = 8$ and 16, respectively, where $J_2$ is negligible; it accounts very well for $T\chi(T)$ in Fig.~\ref{fig8} 
between $T = 0.1$ and 1. Spin excitations of order $1/U$ 
are increasingly well separated from charge excitations of order $U$.

\begin{figure}[]
        \includegraphics[width=\columnwidth]{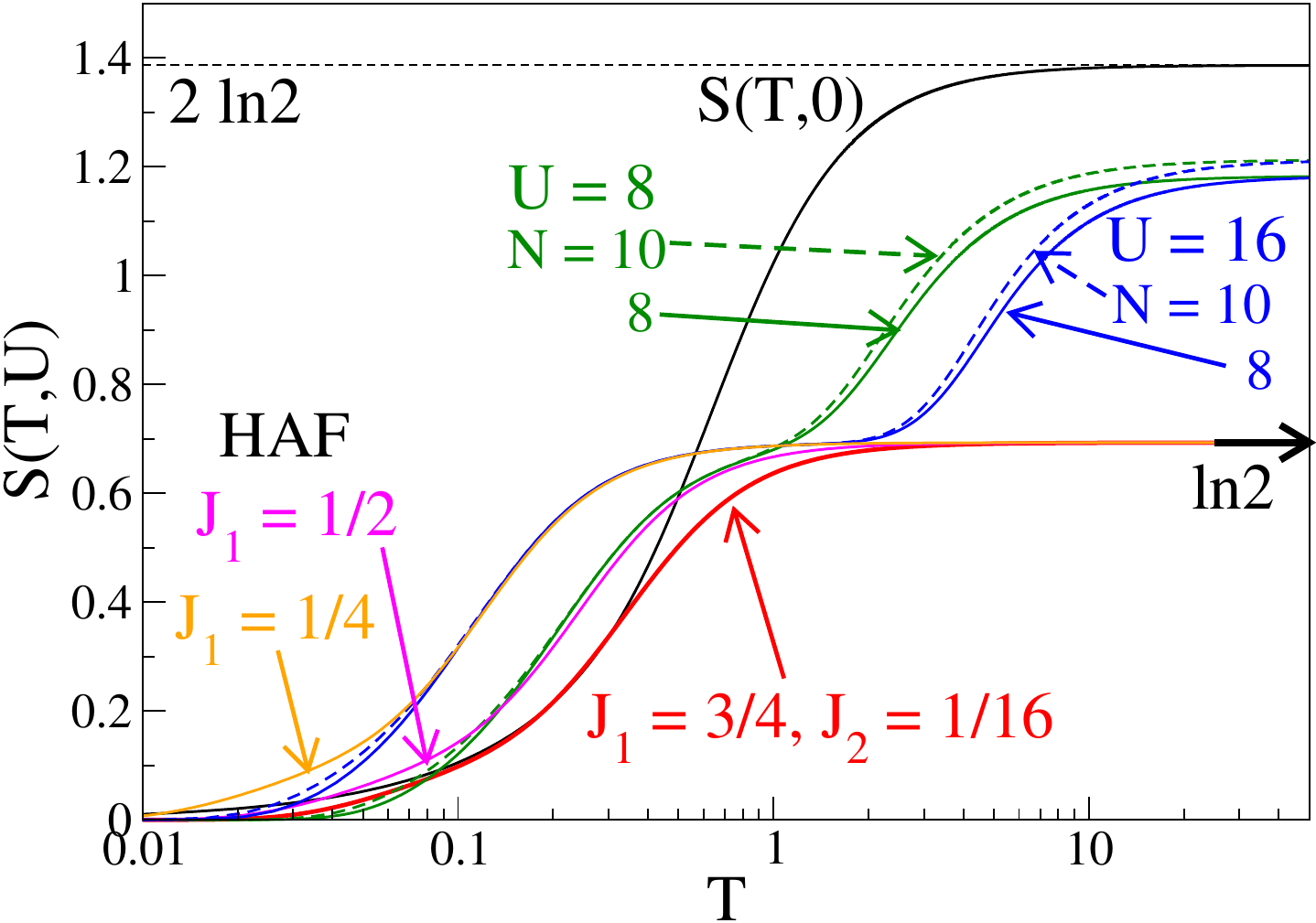}
\caption{\label{fig9}
Calculated $S(T,U,N)$ at $U = 8$ and 10 at system sizes 8 and 10; the band limit is $S(T,0)$. The spin $S(T)$ are HAFs with $J_1 = 1/4$ and 1/2, 
	and the $J_1-J_2$ model with $J_1 = 3/4$, $J_2 = 1/16$.}
\end{figure}

Charge degrees of freedom contribute directly to the entropy of Hubbard models. The contrast with spin chains 
is equally prominent. Fig.~\ref{fig9} shows the band limit and exact $S(T,U,N)$ curves with $U = 8$ and 16 at system sizes 
8 and 10. Finite size effects are again seen at both low and high $T$. Convergence to $2\ln{2}$ requires the scaled entropy discussed 
in Section~\ref{sec2}. The HAF with $J_1 = 1/2$ and 1/4, respectively, fits $S(T,U,N)$ in the intervals $0.12 < T < 0.9$ at $U = 8$  
and $0.08 < T < 1.4$ at $U = 16$. The spin chain gives the thermodynamic limit to much lower $T$ when combined with DMRG for $N > 24$.

The band entropy $S(T,0)$ in Fig.~\ref{fig9} is surprisingly close to the spin entropy of the $J_1 = 3/4$, $J_2 = 1/4$ 
chain up to $T \sim 0.3$. Finite $U$ initially shifts spin excitations to lower energy and charge excitations to higher energy. 
Offsetting entropy changes are a qualitative explanation for similar $S(T,U)$ at low $T$ for $U = 0$ and 4. By inspection, 
the derivative $S^\prime(T,U,N)$ at $U = 8$ or 16 has two maxima separated by a minimum that is close to zero around $T \sim 1$ 
and widens with increasing $U$. Hence $C(T,U,N) = TS^\prime(T,U,N)$ also has two maxima.

Spin-charge separation at $U = 8$ was highlighted in the exact thermodynamics of the half-filled 
Hubbard model~\cite{juttner98}. The exact $\chi(T,8)$ and $C(T,8)$ are used in Ref.~\citenum{sirker2007} to 
demonstrate the impressive accuracy of TMRG down to $T \sim 0.1$ and to emphasize that the method is 
applicable to models whose exact thermodynamics is not known. We make exactly the same 
case for ED/DMRG and the low $T$ thermodynamics that is less or not accessible to TMRG.

\subsection{\label{sec3c}Quarter-filled models}
\begin{figure}
        \includegraphics[width=\columnwidth]{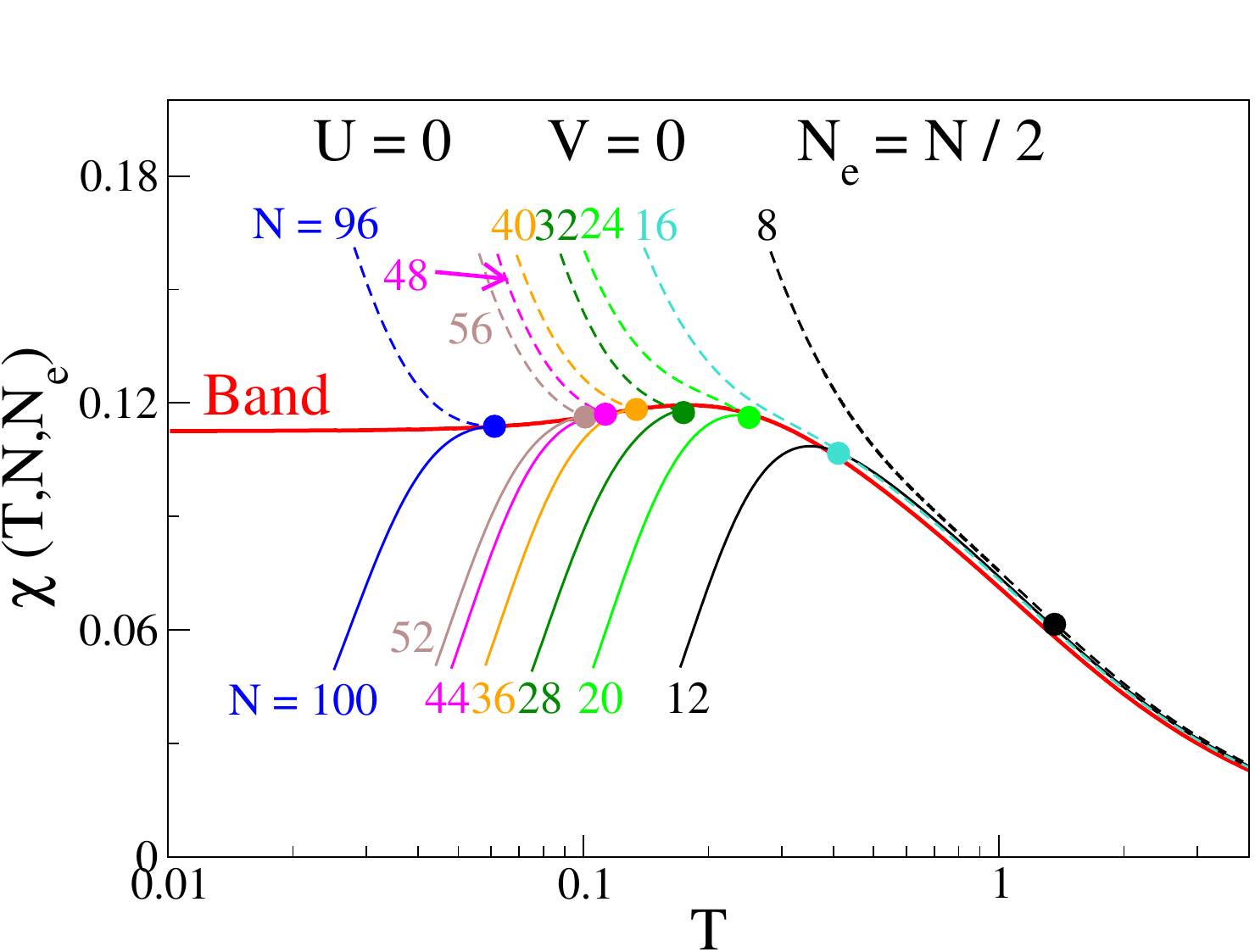}
\caption{\label{fig10}
Susceptibility $\chi(T,N,N_e)$ of quarter-filled models with $U = V = 0$ in Eq.~\ref{eq:EHM_ham}. The points are the merging 
	of $N_e = 4p$, $4p + 2$ lines and define the thermodynamic limit as discussed in Fig.~\ref{fig1} for half-filled models. 
	The red line is the exact band limit $\chi(T,1/2)$ for noninteracting fermions. } 
\end{figure}
We consider Eq.~\ref{eq:EHM_ham} with $N_e = N/2$ electrons or holes. Exact thermodynamics are readily obtained in the 
band limit. Shiba~\cite{shiba72} reported the $T = 0$ susceptibility $\chi(0,U,n)$ at any $U$ or filling $n$ and $V = 0$.

Fig.~\ref{fig10} compares the band limit $\chi(T)$ to $\chi(T,N)$ at $n = 1/2$ using the notation in Fig.~\ref{fig1}. 
The $T = 0$ susceptibility is $\sqrt{2}$ larger and the $\chi(T)$ maximum is at lower $T$. Finite systems with $N_e = 4p$ 
and $4p + 2$ converge from above and below, respectively, for $N = 2N_e$. ED up to $N = 16$ deviates upward at high $T$ 
where the thermodynamic limit of the fraction of singly occupied sites is $\rho_1(n) = n(1 - n/2) = 3/8$. The fraction $\rho_1(N,N/2)$ 
follows from Eq.~\ref{eq:finitebasis}
\begin{equation}
{\rho}_{1}(N,N/2) = \sum_{q=0}^{{N} / {4}}  \left(\frac {N-4q} {2(N-q)} \right)   
	 \frac{W(q,N,N/2)} {W \left(N,N/2 \right)}.
\label{eq:density_quarfill}
\end{equation}
We obtain $\rho_1(N,N/2) =  0.4085$ and 0.4063 at $N = 12$ and 16. DMRG calculations lead to $\chi(T,N)$ curves $N$, $N + 4$ 
that merge at points that define the thermodynamic limit as discussed in connection with half-filled systems. 
The accuracy at quarter filling is quite comparable to half filling. 

\begin{figure}[]
	\includegraphics[width=\columnwidth]{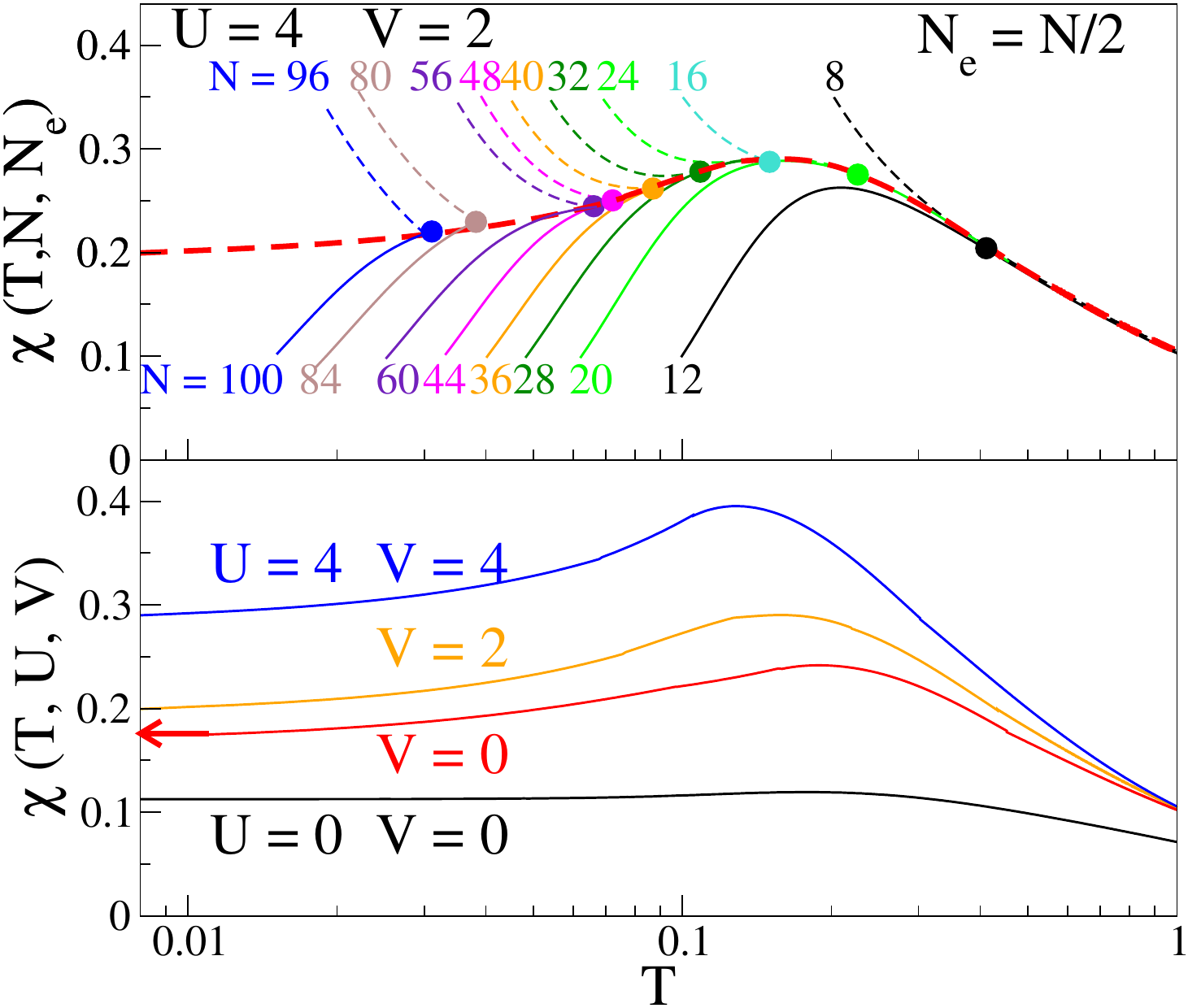}
\caption{\label{fig11}
Upper panel: $\chi(T,N,N_e)$ of quarter-filled models with $U = 4$, $V = 2$ in Eq.~\ref{eq:EHM_ham}. 
	The points are the merging of $N_e = 4p$, $4p + 2$ lines that generate the thermodynamic limit 
	shown a red dashed line. Lower panel: Thermodynamic limit of quarter-filled models at the indicated $U$, $V$. The arrow at $\chi(0)$ is exact~\cite{shiba72}.}
\end{figure}

Fig.~\ref{fig11} has EHM results for $\chi(T,N)$ with $U = 4$, $N_e/N = 1/2$ and $V = 0, 2$ and 4. ED to $N = 12$ is followed 
by DMRG to $N = 100$. The upper panel has $V = 2$ and shows the merging 
points of systems with $N_e = 4p$ and $4p + 2$. The thermodynamic limit is the bold dashed red line. The same construction gives 
the thermodynamic limit of $\chi(T)$ in the lower panel for the indicated $U$, $V$ and the band limit. The arrow marks~\cite{shiba72} the 
exact $\chi(0)$. Increasing either $U$ or $V$ increases the susceptibility by reducing the density of 
doubly occupied diamagnetic sites or by reducing the density of adjacent singly occupied sites that form singlet pairs. 
The large $U$, $V$ limit at quarter filling has $n_r = 1$ at every other site. The system has a singlet ground state, 
weak antiferromagnetic interactions and a charge gap.

To conclude this Section, we comment on arbitrary filling $n$. The DMRG methodology
developed for low-$T$ thermodynamics is robust. Hundreds of low-energy states $R(N)$ are
obtained at $n$ in sectors with $S^Z \le N_e/2$. Convergence of the maximum of $S(T,N)/T$ with increasing $R(N)$ returns the thermodynamic limit of $S'(T) = C(T)/T$ at $T(N)$. Since the ground state of Eq.~\ref{eq:EHM_ham} is a singlet for the parameters of interest, the relevant finite systems must have integral $N$ and even $N_e$. ED for $N_e = 4$ and 6 is minimally required to demonstrate convergence at high $T$, and adding two electrons sets the minimum increase of the system size $N$. We must approximate arbitrary $n$ with commensurate filling. For example, either sublattice of TTF-TCNQ
has $n \sim 0.6$. The smallest system has 6 electrons on 10 sites, suitable for ED,
as is 8 electrons on 13 sites ($n = 0.615$).


\subsection{\label{sec3d}BOW phase}
\begin{figure}[h]
        \includegraphics[width=\columnwidth]{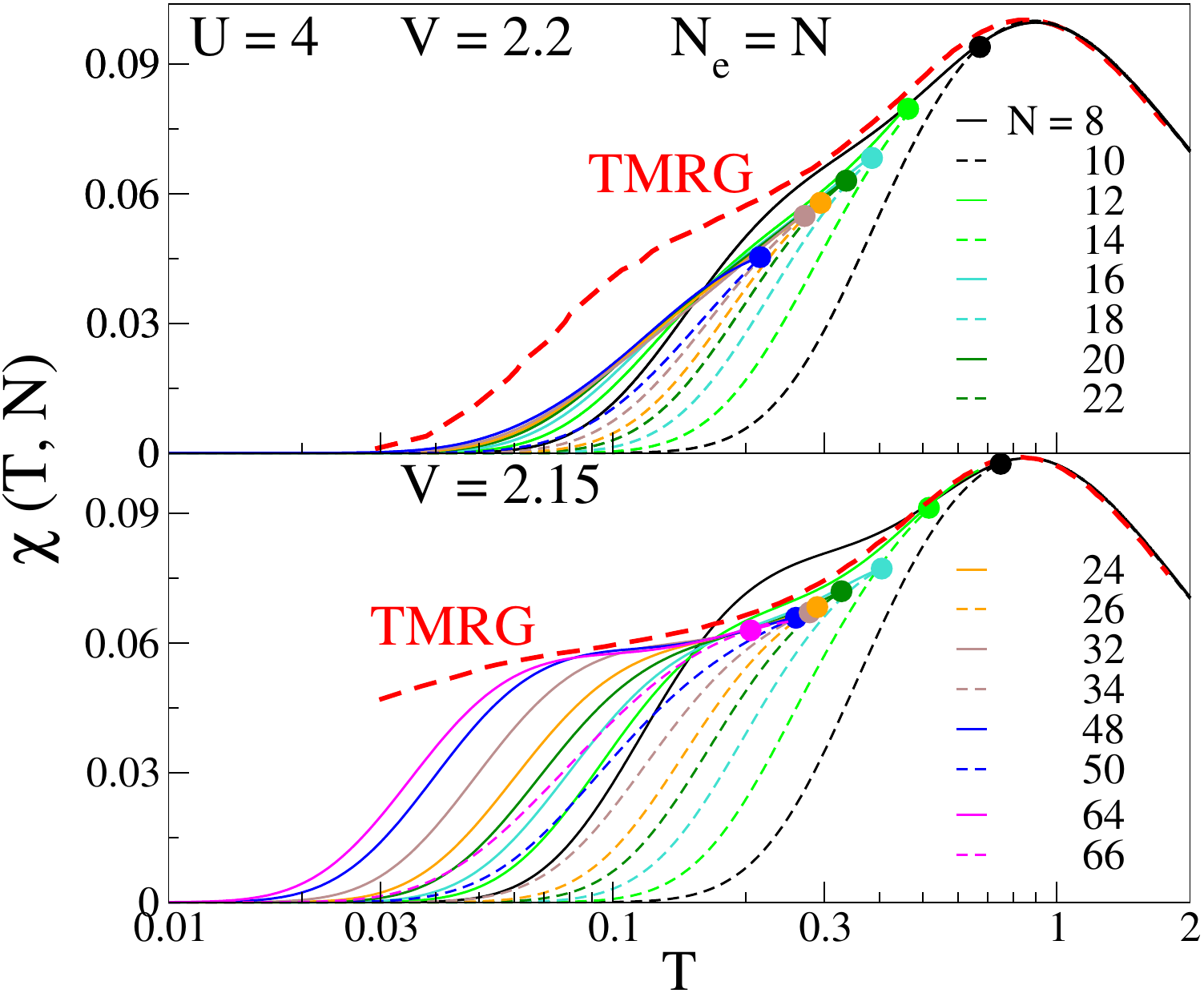}
\caption{\label{fig12}
Susceptibilities $\chi(T,N)$ of half-filled EHM with $U = 4$ and $V = 2.20$ 
	(upper panel) and $V = 2.15$ (lower panel) and merging points $N = 4p$, $4p + 2$. 
	The TMRG curves are from Fig. 13 of Ref.~\citenum{sirker2007}.}
\end{figure}
The quantum phase diagram of the half-filled EHM illustrates competition among $U$, $V$ and $t$. As predicted by Nakamura~\cite{nakamurauv}, intermediate $U$ leads to a narrow 
bond-order-wave (BOW) phase with a finite singlet-triplet gap $E_{ST}$ and doubly degenerate singlet ground state. 
Ongoing studies discuss BOW properties of the EHM and related models~\cite{mkumar2009} with other intersite interactions in Eq.~\ref{eq:EHM_ham}. 
The quantum critical points $V_s$ and $V_c$ at the BOW boundaries are estimated in finite systems using level crossing~\cite{kumaruv}. 
For constant $U < 7$ and $N = 4p$, the EHM has $E_{ST}(4p) = E_\sigma(4p)$ at $V_s(4p) < U/2$, where $E_\sigma(4p)$ 
is the gap between the lowest two singlets, and $E_\sigma(4p) = E_J(4p)$ at $V_c(4p) > U/2$, where $E_J(4p)$ is the gap 
to the lowest singlet with opposite electron-hole symmetry to the ground state. The weak size dependence of the 
critical points allows accurate extrapolation to the thermodynamic limit as discussed~\cite{nomura1992,mkumar2015} 
for the quantum critical point $J_2/J_1 = 0.2241$ of the spin chain, Eq.~\ref{eq:j1j2}.

Glocke et al. reported~\cite{sirker2007} TMRG results for the half-filled EHM at $U = 4$ 
and variable $V$. They find $V_c = 2.165$, fully consistent with level crossing, and note that thermodynamics does not 
determine $V_s = 2.02 \pm 0.06$ accurately. The charge gap $E_J(V)$ vanishes at the boundary $V_c$ between the BOW and the 
charge-density-wave (CDW) phases. Level crossing returns $V_s \sim 1.86$ for the Kosterlitz-Thouless transition where an 
exponentially small $E_{ST}(V)$ opens. We compare below ED/DMRG results for $\chi(T)$ at $U = 4$ and variable $V$ with TMRG results 
in Fig. 13 of Ref.~\citenum{sirker2007}. 

Fig~\ref{fig12} shows $\chi(T,N)$ of gapped models: $V = 2.20$ in the upper panel is just above $V_c$ while $V = 2.15$ 
in the lower panel is in the BOW phase. The points indicate the merging of $\chi(T,N)$ for $N = 4p$, $4p + 2$ using DMRG for $N \geq 14$ 
and ED for $N = 8$, 10 and 12. Here the strong $4p$, $4p + 2$ dependence is due $E_\sigma(4p) = 0$ at $V_1(4p)$, but there is no such degeneracy 
for $4p + 2$. The roles are reversed for antiperiodic boundary conditions~\cite{kumaruv}, when $E_\sigma(4n+2) = 0$ at $V_1(4p+2)$. The thermodynamic 
limit is reached at $N > 1/E_{ST}$ when finite size effects become small compared to $E_{ST}(V)$. The $V = 2.20$ panel shows almost 
converged $\chi(T,4p)$ at low $T$ at system size $4p = 48$; this is an accurate lower bound on $\chi(T)$ at $V = 2.20$. 
The smaller gap in the $V = 2.15$ panel is evident from the $\chi(T,4p)$ size dependence up to $4p = 64$. We estimate that convergence 
to $\chi(T)$ requires $4p \sim 100$.
\begin{figure}[t]
        \includegraphics[width=\columnwidth]{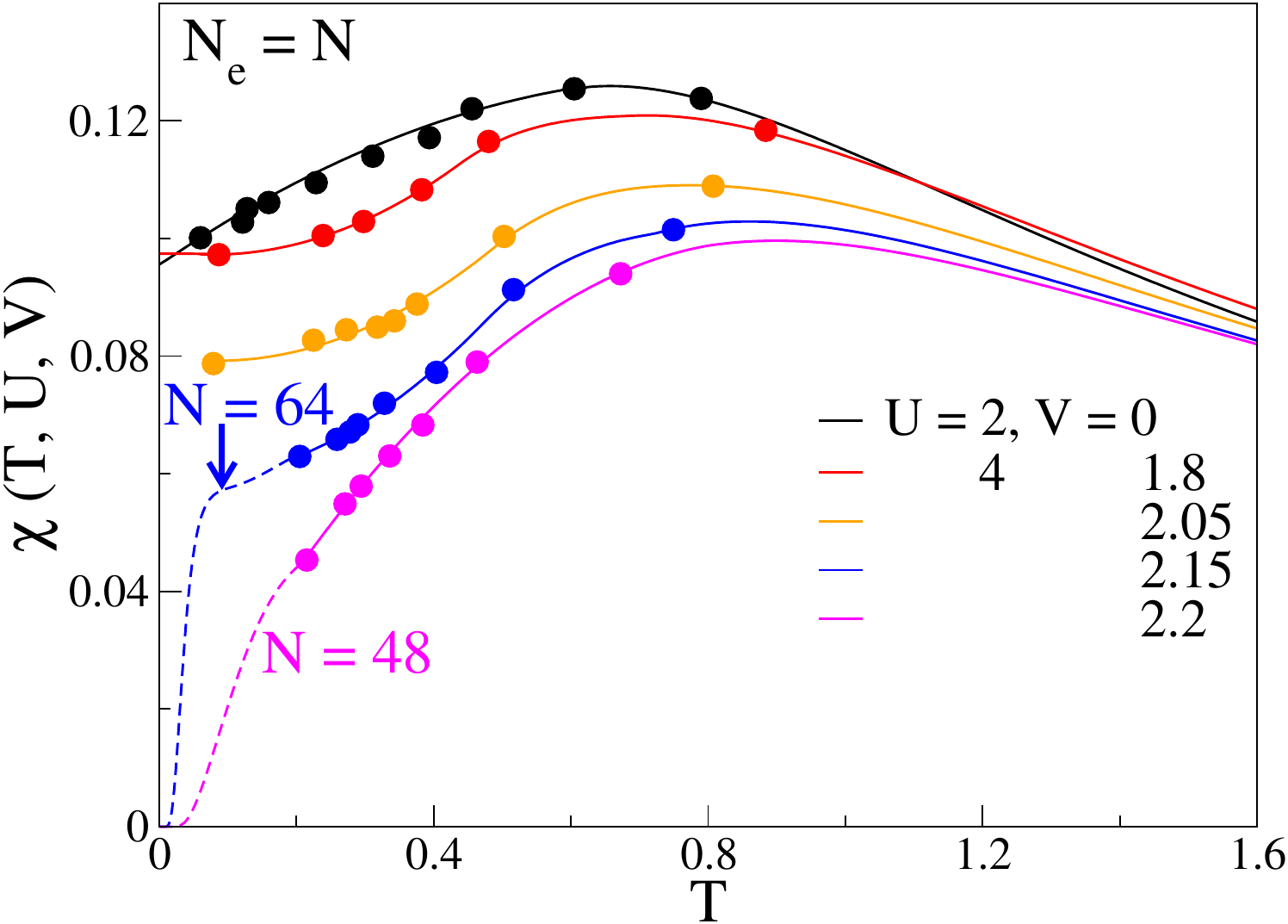}
\caption{\label{fig13}
Thermodynamic limit of $\chi(T,U,V)$ inferred from DMRG calculation with $N = 4p$, $4p + 2$ merging points 
	shown and discussed in the text; ED gives higher $T$. The dashed $V = 2.20$ and 2.15 lines are based 
	on $N = 48$ and 64, respectively. Gapless models have $\chi(0,U,V) > 0$. The small gap for $V = 2.05$ 
	requires larger systems.}
\end{figure}

TMRG substantially overestimates $\chi(T)$ at $V = 2.20$ for $T < 0.8$ because the reported~\cite{sirker2007} $E_{ST} = 0.11$ 
is smaller than $0.182$ obtained by DMRG and $1/N$ extrapolation of $E_{ST}(N)$ up to $N = 100$. TMRG at $V = 2.15$ and the slight 
downturn of $\chi(T)$ is the first indication of finite $E_{ST} = 0.039$, again smaller than 0.065, the extrapolated DMRG gap. 
There is fair agreement for $T > 0.1$ and good agreement in both panels for $T > 0.5$.

We also studied $V = 1.80$ in the gapless phase and $V = 2.05$ in the BOW phase. Agreement with TMRG is good at $T > 0.1$ in both cases. 
The $\chi(T,U,V)$ curves in Fig.~\ref{fig13} are based on ED at high $T$ and the DMRG merging points at low $T$, 
as shown in Fig.~\ref{fig12} for $V = 2.15$ and 2.20, and in Fig.~\ref{fig6} for $U = 2$, $V = 0$. Increasing $V$ 
decreases $\chi(T,4,V)$ by raising the energy of adjacent $n_r = 1$ sites. Since the EHM has some resemblance to a 
Hubbard model with $U_{eff} = U - V$, similar $\chi(T,U,V)$ at $T > 1$ is as expected.

The low $T$ behavior in Fig.~\ref{fig13} are quite different. The $V = 1.80$ results for $\chi(T,N)$ have $4p$, $4p + 2$ patterns 
similar to those in the upper panel of Fig.~\ref{fig6} for $U = 2$, $V = 0$. Convergence occurs at smaller system size 48/50. 
Extrapolation gives finite $\chi(0)$ in gapless systems. As noted above, $N = 48$ is close to the thermodynamic limit at $V = 2.20$ 
while $N = 64$ is a lower bound at $V = 2.15$. The sharp initial increase of $\chi(T)$ is governed by finite $E_{ST}$. Accordingly, 
the tiny gap at $V = 2.05$ presumably leads to an even more abrupt $\chi(T)$ rise in Fig.~\ref{fig13} to the $\chi(T) \sim 0.08$ 
plateau in systems of several hundred sites. Exponentially small $E_{ST}$ at $V$ slightly larger than $V_s$ is not relevant 
to the thermodynamics except at exponentially low $T$.
\begin{figure}[t]
        \includegraphics[width=\columnwidth]{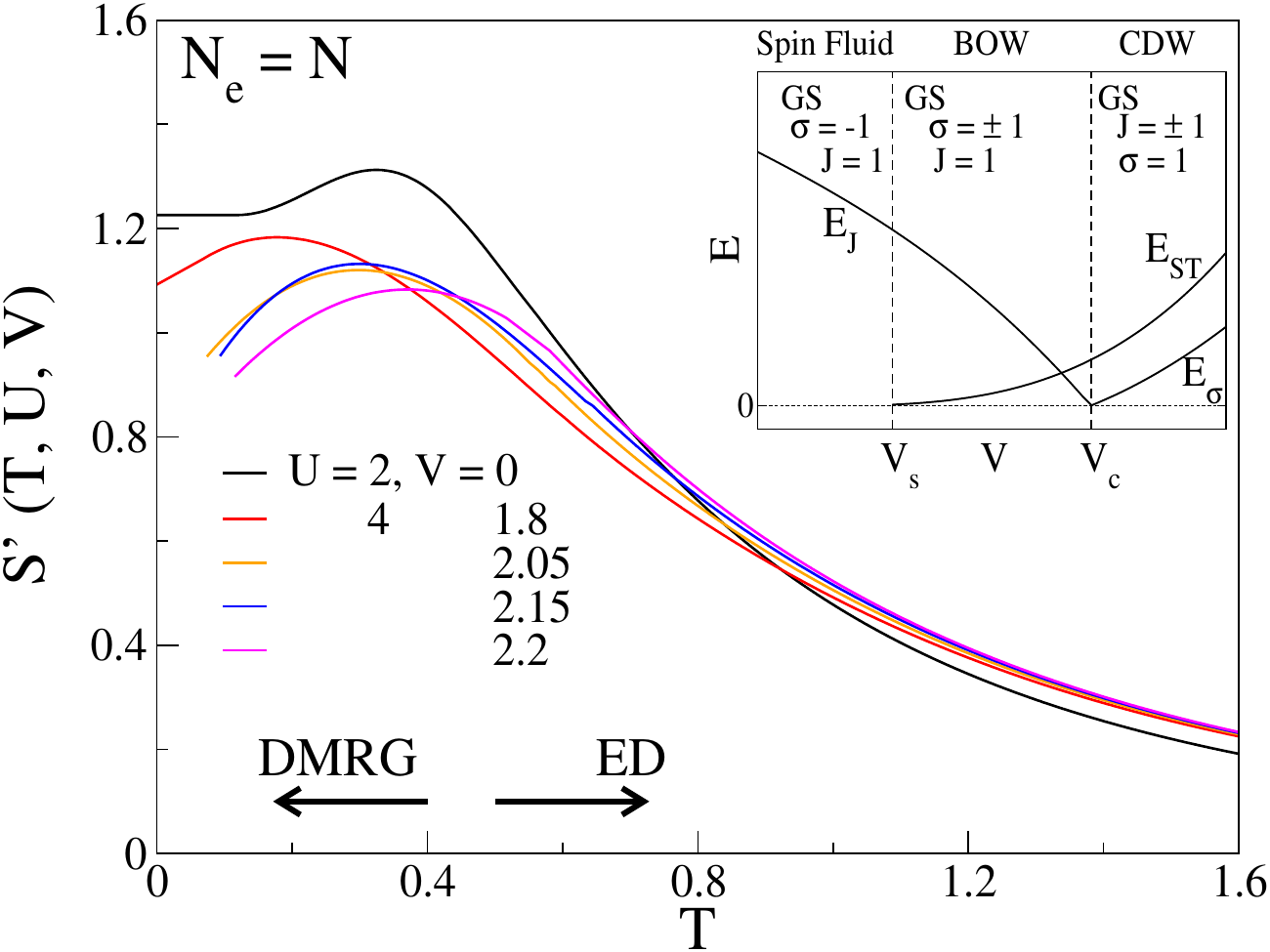}
\caption{\label{fig14}
Thermodynamic limit of entropy derivative $S^\prime(T,U,V) = C(T,U,V)/T$ inferred from DMRG calculations at $T < 0.4$ 
	discussed in the text and ED at higher $T$. The inset shows the BOW boundaries $V_s$ and $V_c$, the ground 
	state degeneracy and the spin and charge gaps.}
\end{figure}

The calculated $\{E(N)\}$ that return $\chi(T,U,V)$ also yield $S(T,U,V)$ as shown in Figs.~\ref{fig5} and ~\ref{fig7} 
using the scaled entropy of finite systems. The entropy at $T \sim 1$ is roughly $75\%$ of the high $T$ limit, consistent 
with a single energy scale in systems with comparable bandwith, $U$ and $V$. The $T > 0.4$ range is given by ED for $N = 10$. 
DMRG for larger $N = 4p + 2$ systems gives $S(T,N)$ down to $T \sim 0.1$. We fit the $T < 0.4$ results for $S(T,U,V)$ 
to a 4th order polynomial and differentiate to obtain the $S^\prime(T,U,V) = C(T,U,V)/T$ curves in Fig.~\ref{fig14} 
for the same systems as in Fig.~\ref{fig13}. The gapless $V = 1.80$ and Hubbard curves are extrapolated to finite $S^\prime(0)$. 
Extrapolation of gapped systems with $S^\prime(0) = 0$ requires larger systems. In contrast to $\chi(T,U,V)$, we find 
similar $S^\prime(T,U,V)$ for the EHMs and clear differences with the Hubbard curve. 

Since the area under $S^\prime(T,U,V)$ curves is $2\ln{2}$ in half-filled systems, the $U = 2$, $V = 0$ 
maximum at $T \sim 0.4$ is offset by lower $S^\prime(T,2,0)$ at $T > 0.8$. The small $S^\prime(T,4,V)$ differences among 
EHM curves are largely compensated by $T \sim 1$. As sketched in the inset, the charge gap is less than $E_{ST}$ on 
either side of $V_c = 2.165$. The ground state is nondegenerate in the gapless phase with $V < V_s$ and doubly 
degenerate in the gapped BOW and CDW phases. The smallest gap governs the low $T$ entropy while the susceptibility 
depends on the spin gap. We understand why there are spin gaps at $V = 2.20$ and $2.15$ in Fig.~\ref{fig13} but do not 
see a charge gap in Fig.~\ref{fig14} at system sizes 48 and 64, respectively, for $V = 2.20$ and 2.15.

\section{\label{sec4} Discussion}

We have applied the ED/DMRG method to the low $T$ thermodynamics of representative 1D fermionic models. The method is general. 
The diagonalization of the superblock is the most time-intensive step at each cycle of DMRG calculations. The 
computational cost for obtaining the lowest few eigenvalues goes as O($m^3$) where $m$ (here $500$) is 
number of eigenvectors of the density matrix corresponding to the largest eigenvalues kept in the calculation. 
The time increases linearly with system size since the dimension of the superblock matrix in Section~\ref{sec2b} is kept 
constant. The system size is limited by the numerical accuracy required for the dense low-energy spectrum $\{E(N)\}$, 
which is model dependent. Hence the lowest accessible $T$ depends on the model as well as 
on effort. It is in the range $0.01 < T/t < 0.05$ for the illustrative systems in Section~\ref{sec3} and 
extrapolation to lower $T$ is often possible.  

Fermionic models are more complex than spin chains. The energy spectrum $\{E(N,N_e)\}$ of fermionic systems depends on the 
filling $n = N_e/N$ as well as the system size in models with $W(N,N_e)$ states. The scaled partition function 
in Section~\ref{sec2a} increases $W(N,N_e)$ by the factor $\lambda(N,n)$ in Eq.~\ref{eq:scale_factor} that 
involves counting and holds for arbitrary interactions. As noted in Section~\ref{sec2}, ED/DMRG is not exact at 
high $T$ due to the size dependence of the basis. Its accuracy is shown by comparisons to the band limit or 
to exact $T = 0$ results at half and quarter filling. The method is limited to models with a single energy scale 
and is best suited for commensurate $n = 1$ or 1/2.

The ED/DMRG method is advantageous for several reasons. First, finite $T$ complements the analysis of 
important models whose $T = 0$ properties such as the ground state, elementary excitations and correlation 
functions are already known, and known exactly in a few notable cases. In particular, the $T > 0 $ analysis 
takes into account whether the model is gapped or gapless. 

Second, low $T$ thermodynamics are obtained by independent calculations on increasingly large systems. We 
have exploited systematic variations with $N$, the convergence to the thermodynamic limit from above or below with increasing $T$, and the convergence of the maximum $T(N)$ of
$S(T,N)/T$ with increasing cutoff $E_C(N)$. Explicit results for finite systems provide some clues about the largely unknown 
properties of correlated excited states. Extrapolations based on systems size have been widely used from the outset.

Third and most importantly, the numerical method is generally applicable to 1D models and can incorporate 
electron-phonon coupling or other interactions. It holds for dimerized chains with alternating electron transfer $t_1$, $t_2$ or 
site energies $\varepsilon_1$, $\varepsilon_2$ or both. The thermodynamic limit is accessible in models with 
Peierls or spin Peierls transitions~\cite{sudipsp2020} when the $T = 0$ gap is large compared to finite size gaps. The 
method is suitable for other thermodynamic properties and for static correlation functions.

\section*{Acknowledgments}

MK thanks Professor Sanjay Singh for the hospitality during the IIT(BHU) visit, and also thanks SERB for financial support through 
grant sanction number CRG/2020/000754. SKS thanks DST-INSPIRE for financial support.


\section*{Appendix}

The ground state $E_1(N,N_e)$ is nondegenerate for $N = N_e = 4p + 2$ in Eq.~\ref{eq:EHM_ham} with $U = V = 0$.
The $S^Z = 0$ spectrum starts with 8-fold degenerate $E_2(N) = 4\sin\pi/N$. The 16-fold degenerate $E_9 = 2(\sin\pi/N + \sin3\pi/N)$
and 18-fold degenerate $E_{26} = 8\sin\pi/N$ are closely spaced. The DMRG entries in Table~\ref{taba1} are the lowest and highest
excitations $p$ and $p^\prime$ in the indicated range. Similar accuracy is found at $N = 26$ and $50$ for $E_2(N)$. The accuracy decreases
at higher energy. As discussed in Section~\ref{sec2c}, the spectrum converges to the thermodynamic limit at and slightly above $T(N)$, and $T(26) = 0.34$,
$T(50) = 0.26$ are comparable to the finite size gaps.

	\begin{table}[h]
  \caption{Exact and DMRG excitations $E_p(N)$ of half-filled bands in the $S^Z = 0$ sector of Eq.~\ref{eq:EHM_ham} with
	$U = V = 0$ at $N = 26$ and $50$. 
	The ground state is $E_1(N) = 0$.}
	\label{taba1}
  \begin{tabular}{llll}
    \hline \hline
   $N = 26$  & Exact  &  DMRG, $p$  & DMRG, $p^\prime$     \\
	  $p$ to $p^\prime$  &  &  &  \\
	  \hline
        2 - 9 &  0.4822 &      0.4842 (0.4$\%$) &     0.4984 (3.3$\%$) \\
        10 - 25 &       0.9503 &       0.9602 (1.0$\%$) &    1.0141 (6.3$\%$) \\
        26 - 43 &       0.9643 &       1.0147 (5.0$\%$) &     1.0465 (7.9$\%$) \\
        44 - 59 &       1.3772 &        1.4258 (3.4$\%$) &     1.4718 (6.4$\%$) \\
        60 - 67 &       1.4184 &       1.4733 (3.7$\%$) &     1.4867 (4.6$\%$) \\
        68 - 139 &      1.4324 &       1.4869 (3.7$\%$) &     1.7502 (18$\%$) \\  
	  \hline 
                $N = 50$  \\ 
		\hline
        2 - 9 & 0.2512 &       0.2547 (1.4$\%$) &     0.2859 (12$\%$) \\
        10 - 25 &       0.5003  &       0.5308  (5.7$\%$) &     0.6018 (17$\%$) \\
        26 - 43 &       0.5023  &       0.6032  (17$\%$) &      0.7858 (36$\%$) \\
        44 - 59 &       0.7436  &       0.8442  (12$\%$) &      0.9224 (19$\%$) \\
        60 - 67 &       0.7495  &       0.9255  (19$\%$) &      0.9623 (22$\%$) \\
        68 - 139 &      0.7515  &       0.9644  (22$\%$) &      1.311 (43$\%$) \\
    \hline
  \end{tabular}
\end{table}

The $N = N_e = 4p$ ground state is 4-fold degenerate in the $S^Z = 0$ sector. The DMRG excitation $E_4$ is $2 \times 10^{-4}$ at $N = 24$
and $3 \times 10^{-3}$ at $N = 48$. The exact excitation $E_5(N) = 2\sin2\pi/N$ is 16-fold degenerate, as is $E_{21} = 2\sin4\pi/N$, while
the double e-h excitation $E_{37} = 4\sin2\pi/N$ is 6-fold degenerate in Table~\ref{taba2}. $T(24) = 0.34$ and $T(48) = 0.28$ are again comparable to the finite size gaps.

	\begin{table}[h]
  \caption{Exact and DMRG excitations $E_p(N)$ of half-filled bands in the $S^Z = 0$ sector at $N = 24$ and 48. The ground state
        is $E_1(N) = 0$.}
  \label{taba2}
  \begin{tabular}{llll}
    \hline
    \hline
$N = 24$ & Exact  &  DMRG, $p$  & DMRG, $p^\prime$     \\
 $p$ to $p^\prime$ &  &  &  \\
	  \hline
        5 - 20 &        0.5176 &  0.5194 (0.4$\%$) &       0.5428 (4.4$\%$)\\ 
        21 - 36 &       1.0 &  1.0557 (5.3$\%$) &       1.0622 (5.9$\%$) \\
        37 - 84 &       1.0353 &  1.0627 (2.6$\%$) &       1.3392 (23$\%$) \\
        85 - 150 & 1.5176 &       1.6121 (5.9$\%$) &       1.7526 (13$\%$) \\ 
	  \hline
                $N = 48$  \\ 
		\hline
        5 - 20 &        0.2611 &  0.2644  (1.3$\%$) &       0.3111 (16$\%$) \\
        21 - 36 &       0.5176 &  0.5506  (6.0$\%$) &  0.5695 (9.1$\%$) \\
        37 - 84 &       0.5221 &  0.5738  (9.0$\%$) & 0.7463 (30$\%$) \\
        85 - 150 & 0.7654 &       0.8725 (12$\%$) & 1.0635 (28$\%$) \\
    \hline
  \end{tabular}
\end{table}
%


The excitations in quarter-filled bands are shown in Table~\ref{taba3}.
The Fermi level at $k_F = \pm \pi/4$ has 2 or 4 electrons when $N_e = N/2$ is $4p$ or $4p + 2$.
The $4p$, $S^Z = 0$ ground state is 4-fold degenerate while the $4p + 2$ ground state is nondegenerate. The degeneracy of exact excitations
is lower than at $N = N_e$ because $\varepsilon(k)$ is not symmetric about $k_F = \pm \pi/4$.


%
\begin{table}[h]
  \caption{Exact and DMRG excitations $E_p(N)$ of quarter-filled bands in the $S^Z = 0$ sector at $N = 64$ and 68. We have $E_1(N) = 0$,
        $E_4(64) = 10^{-3}$ and $T(68) = 0.12$ is close to the finite size gap. }
  \label{taba3}
  \begin{tabular}{llll}
    \hline
    \hline
   $N = 64$  & Exact  &  DMRG, $p$  & DMRG, $p^\prime$     \\
	  $p$ to $p^\prime$ & & & \\
	  \hline
        5 - 12 &        0.1318  &       0.1327  (0.7$\%$) &     0.1469  (10$\%$) \\
        13 - 20 &       0.1454  &       0.1594  (8.8$\%$) &     0.1778  (18$\%$) \\
        21 - 28 &       0.2487  &       0.2827  (12$\%$) &      0.2912  (15$\%$) \\
        29 - 32 &       0.2636  &       0.2916  (9.6$\%$) &     0.2928  (10$\%$) \\
        33 - 72 &       0.2772  &       0.2928  (5.3$\%$) &     0.4420  (37$\%$) \\
         73 - 76 &       0.2908  &       0.4442  (35$\%$) &      0.4594  (37$\%$) \\
         77 - 84 &       0.3030  &       0.4629  (35$\%$) &      0.4687  (35$\%$) \\
         85 - 88 &       0.3496  &       0.4694  (26$\%$) &      0.4728  (26$\%$) \\
         89 - 96 &       0.3805  &       0.4732  (20$\%$) &      0.4784  (20$\%$) \\  
	  \hline
                $N = 68$  \\ 
		\hline
        2 - 9 &  0.1306 &      0.1318 (0.9$\%$) &     0.1431 (8.8$\%$) \\
        10 - 17 &       0.2486  &       0.2629  (5.4$\%$) &     0.2774  (10$\%$) \\
        18 - 35 &       0.2612  &       0.2833  (7.8$\%$) &     0.3426  (24$\%$) \\
        36 - 43 &       0.2727  &       0.3470  (21$\%$) &      0.4060  (33$\%$) \\
        44 - 51 &       0.3530  &       0.4061  (13$\%$) &      0.4128  (15$\%$) \\
         52 - 91 &       0.3792  &       0.4148  (8.6$\%$) &     0.5452  (30$\%$) \\
         92 - 99 &       0.3907  &       0.5483  (28$\%$) &      0.5752  (32$\%$) \\
        100 - 107 &     0.3918  &       0.5819  (33$\%$) &      0.5892  (34$\%$) \\
	  \hline
  \end{tabular}
\end{table}

\pagebreak
\newpage


\end{document}